\documentclass[twocolumn,tighten,times]{aastex62}

\usepackage{mathrsfs}

\newcommand{\hb}{\ifmmode {\rm H\beta} \else H$\beta$\fi}
\newcommand{\mgii}{\ifmmode {\rm Mg\ II} \else Mg {\sc ii}\fi}
\newcommand{\feii}{\ifmmode {\rm Fe\ II} \else Fe {\sc ii}\fi}
\newcommand{\heii}{\ifmmode {\rm He\ II} \else He {\sc ii}\fi}
\newcommand{\oiii}{\ifmmode {\rm [O\ III]} \else [O {\sc iii}]\fi}
\newcommand{\mbh}{\ifmmode {M_{\bullet}} \else $M_{\bullet}$\fi}
\newcommand{\fblr}{\ifmmode {f_{\rm BLR}} \else $f_{\rm BLR}$\fi}
\newcommand{\rhb}{\ifmmode {R_{\rm H\beta}} \else $R_{\rm H\beta}$\fi}
\newcommand{\rblr}{\ifmmode {R_{\rm BLR}} \else $R_{\rm BLR}$\fi}
\newcommand{\taublr}{\ifmmode {\tau_{\rm BLR}} \else $\tau_{\rm BLR}$\fi}
\newcommand{\tauhb}{\ifmmode {\tau_{\rm H\beta}} \else $\tau_{\rm H\beta}$\fi}
\newcommand{\dotm}{\ifmmode {\dot{\mathscr{M}}} \else $\dot{\mathscr{M}}$\fi}

\def\kms{{\rm km~s^{-1}}}

\begin{document}

\title{MONITORING AGNS WITH \hb\ ASYMMETRY. I. FIRST RESULTS: \\
VELOCITY-RESOLVED REVERBERATION MAPPING}

\author{Pu Du}
\affiliation{Key Laboratory for Particle Astrophysics, Institute of High Energy Physics,
Chinese Academy of Sciences, 19B Yuquan Road, Beijing 100049, China;\\ \url{dupu@ihep.ac.cn}, \url{wangjm@ihep.ac.cn}}

\author{Michael S. Brotherton}
\affiliation{Department of Physics and Astronomy, University of Wyoming, Laramie, WY 82071, USA; \url{mbrother@uwyo.edu}}

\author{Kai Wang}
\affiliation{Key Laboratory for Particle Astrophysics, Institute of High Energy Physics,
Chinese Academy of Sciences, 19B Yuquan Road, Beijing 100049, China;\\ \url{dupu@ihep.ac.cn}, \url{wangjm@ihep.ac.cn}}

\author{Zheng-Peng Huang}
\affiliation{Key Laboratory for Particle Astrophysics, Institute of High Energy Physics,
Chinese Academy of Sciences, 19B Yuquan Road, Beijing 100049, China;\\ \url{dupu@ihep.ac.cn}, \url{wangjm@ihep.ac.cn}}

\author{Chen Hu}
\affiliation{Key Laboratory for Particle Astrophysics, Institute of High Energy Physics,
Chinese Academy of Sciences, 19B Yuquan Road, Beijing 100049, China;\\ \url{dupu@ihep.ac.cn}, \url{wangjm@ihep.ac.cn}}

\author{David H. Kasper}
\affiliation{Department of Physics and Astronomy, University of Wyoming, Laramie, WY 82071, USA; \url{mbrother@uwyo.edu}}

\author{William T. Chick}
\affiliation{Department of Physics and Astronomy, University of Wyoming, Laramie, WY 82071, USA; \url{mbrother@uwyo.edu}}

\author{My L. Nguyen}
\affiliation{Department of Physics and Astronomy, University of Wyoming, Laramie, WY 82071, USA; \url{mbrother@uwyo.edu}}

\author{Jaya Maithil}
\affiliation{Department of Physics and Astronomy, University of Wyoming, Laramie, WY 82071, USA; \url{mbrother@uwyo.edu}}

\author{Derek Hand}
\affiliation{Department of Physics and Astronomy, University of Wyoming, Laramie, WY 82071, USA; \url{mbrother@uwyo.edu}}

\author{Yan-Rong Li}
\affiliation{Key Laboratory for Particle Astrophysics, Institute of High Energy Physics,
Chinese Academy of Sciences, 19B Yuquan Road, Beijing 100049, China;\\ \url{dupu@ihep.ac.cn}, \url{wangjm@ihep.ac.cn}}

\author{Luis C. Ho}
\affiliation{Kavli Institute for Astronomy and Astrophysics, Peking University, Beijing 100871, China}
\affiliation{Department of Astronomy, School of Physics, Peking University, Beijing 100871, China}

\author{Jin-Ming Bai}
\affiliation{Yunnan Observatories, Chinese Academy of Sciences, Kunming 650011, China}

\author{Wei-Hao Bian}
\affiliation{Physics Department, Nanjing Normal University, Nanjing 210097, China}

\author{Jian-Min Wang}
\altaffiliation{PI of the MAHA Project}
\affiliation{Key Laboratory for Particle Astrophysics, Institute of High Energy Physics,
Chinese Academy of Sciences, 19B Yuquan Road, Beijing 100049, China;\\ \url{dupu@ihep.ac.cn}, \url{wangjm@ihep.ac.cn}}
\affiliation{National Astronomical Observatories of China, Chinese Academy of Sciences, 20A Datun Road, Beijing 100020, China}
\affiliation{School of Astronomy and Space Science, University of Chinese Academy of Sciences, 19A Yuquan Road, Beijing 100049, China}

\collaboration{(MAHA Collaboration)}

\journalinfo{To appear in {\it The Astrophysical Journal}.}

\begin{abstract} 
We have started a long-term reverberation mapping project using the Wyoming Infrared Observatory 
2.3 meter telescope titled ``Monitoring AGNs with \hb\ Asymmetry" (MAHA).
The motivations of the project are to explore the geometry and kinematics of the gas responsible 
for complex \hb\ emission-line profiles, ideally leading to an understanding
of the structures and origins of the broad-line region (BLR).
Furthermore, such a project provides the opportunity to
search for evidence of close binary supermassive black holes.
We describe MAHA and report initial results from our first campaign,
from December 2016 to May 2017, highlighting velocity-resolved time lags for four AGNs
with asymmetric H$\beta$ lines.  We find that 3C~120, Ark~120, and Mrk~6  display complex features different 
from the simple signatures expected for pure outflow, inflow, or a Keplerian disk.  
While three of the objects have been previously reverberation mapped, including 
velocity-resolved time lags in the cases of 3C~120 and Mrk~6, 
we report a time lag and corresponding black hole mass measurement for 
SBS~1518$+$593 for the first time.  Furthermore, SBS~1518$+$593, the least asymmetric of 
the four, does show velocity-resolved time lags characteristic of a Keplerian disk or virialized motion more generally.
Also, the velocity-resolved time lags of 3C~120 have significantly changed
since previously observed, indicating an evolution of its BLR structure.
Future analyses of the data for these objects and others in MAHA 
will explore the full diversity of H$\beta$ lines and the physics of AGN BLRs.
\end{abstract}


\section{Introduction}
The most prominent features in the UV and optical spectra of luminous active galactic
nuclei (AGNs) and quasars are the broad emission lines (BELs), with velocity widths 
ranging from $\sim10^3\,\kms$ to  $\sim10^4\,\kms$ \cite[e.g.,][and references
therein]{schmidt1963, Osterbrock1986, boroson1992, sulentic2000, shen2011}. 
The spectra generally show similarities from local
Seyfert galaxies all the way to $z\gtrsim 7$ quasars 
\cite[e.g.,][]{francis1995,vandenBerk2001,Mortlock2011,Banados2018}.
This suggests that the broad-line regions (BLRs) of AGNs have similarities 
from object to object over cosmological time. The similarities
lead to recognition of a common formation mechanism of the BLRs
in most AGNs and quasars, 
and also to the importance of investigating nearby AGNs for insight into high$-z$ quasars.
However, we so far possess insufficient understanding of the BLRs and their physics.

Reverberation mapping (RM) campaigns \citep[e.g.,][]{peterson1993, peterson1998, peterson2002, peterson2004,
kaspi2000, kaspi2007, bentz2008, bentz2009, denney2009, barth2011, barth2013,
barth2015, rafter2011, rafter2013, du2014, du2015, du2016V, du2018, wang2014,
shen2016, fausnaugh2017, grier2012, grier2017} provide measurements of
BEL lags ($\tau_{\rm BLR}$) with respect to the varying ionizing continuum
as originally suggested by \cite{bahcall1972} and \cite{blandford1982}.
The time lag multiplied by the speed of light ($c$) provides the size scale of the 
BLR. In individual objects, different lines echoing at different distances are consistent with the same virial black hole mass
\begin{equation}
\label{eqn:mass}
\mbh = \fblr\frac{\rblr \Delta V^2}{G},
\end{equation}
indicating Keplerian motion, 
where $G$ is the gravitational constant,
$\rblr=c\taublr$ is the emissivity-weighted radius of the BLR, 
$\Delta V$ is either full-width-half-maximum (FWHM) or velocity 
dispersion ($\sigma_{\rm line}$) 
of mean spectra or root-mean-square (rms) spectra 
\citep{peterson1999,wandel1999}, 
and $\fblr$ is the virial factor determined by
the geometry and kinematics of the BLR \citep[e.g.,][]{peterson2004}. 
In past decades, RM established a tight correlation between time lags 
and luminosity based on a heterogeneous AGN sample containing mostly 
sub-Eddington sources \citep{kaspi2000,bentz2013}. More recently, however, new RM campaigns have discovered that lags are substantially shorter for super-Eddington AGNs 
than their sub-Eddington counterparts of the same luminosity  \citep{du2015,du2016V,du2018}.
Beyond just determining 
a single global time lag, high-quality reverberation mapping can provide time lags as a function of velocity.  Such velocity-resolved time lags are diagnostic of the BLR
structure itself, and these and other advanced analysis products such as velocity-delay maps and dynamical models commonly find that  the BLR is comprised of a thick Keplerian 
disk \citep[e.g.,][]{bentz2008, denney2010, grier2013, pancoast2014b,du2016VI,grier2017vdm}\footnote{There 
are several other lines of evidence indicate that the typical BLR has the structure and dynamics 
of a flattened disk with emission-line gas following Keplerian motion (see \citealt{gaskell2009} for 
a review of the BLR).  For instance, \hb\ velocity widths are systematically smaller in more 
face-on systems as determined using orientation-dependent radio properties \citep{wills1986}, 
and spectropolarimetry of type 1 AGNs also supports a flattened, disk-like 
geometry \citep{smith2005,baldi2016,savic2018}.}.
This flattened disk of BLRs as the major ingredient could be common in AGNs and quasars.
The BLR, however, appears significantly more complex in many objects.

The BEL profiles (especially those of the \hb\ line) of most AGNs
are roughly symmetric (e.g., Gaussian or Lorentzian-like), but there are
still a large number of objects showing asymmetric (redward, blueward or even
double-peaked) profiles both for low and high-ionization lines
\citep[e.g.,][]{boroson1992, eracleous1994, brotherton1996, peterson1997}.
Similarly, \hb\ usually shows a line peak close to systemic velocity, but 
extreme outliers exist \citep{eracleous2012, shen2016V}.  Some net radial motion and/or 
opacity effects would seem required to explain BLRs displaying extreme profiles.
The BELs with the extreme profiles may indicate special
geometry and kinematics of their BLRs, such as super-fast or ultra-strong inflow
or outflow, or even abnormal nuclear environments (e.g., absorption or dust).
The asymmetries of H$\beta$ profiles significantly
correlate with the ratio of \feii\ to H$\beta$ \citep{boroson1992}, and
with \oiii\, luminosities \citep{wang2017a}.
High-precision RM has shown indications of 
infall and (more rarely) outflow in \hb-emitting gas, sometimes in the presence of a 
Keplerian disk component as well \citep[e.g.,][]{grier2017vdm}.  Some objects with excellent 
RM data sets have defied simple explanation, such as Mrk 6 \citep{doroshenko2012,grier2013}. 
Mrk 6 is also noteworthy for its complex \hb\ profile, which possesses a strongly blueshifted 
peak in addition to a lower velocity peak in most epochs, and red asymmetric tail.
More detailed investigations of such AGNs are needed to provide a deeper understanding 
of the full diversity of the structure and kinematics of the BLR, and perhaps clues to 
its origin as well.  The latter has been a controversial subject that is not yet resolved, 
although there are many proposals \citep{murray1997, czerny2011, elvis2017, wang2017a, baskin2018}.

\begin{deluxetable}{lllccc}
\tablecolumns{6}
\tablewidth{\textwidth}
\setlength{\tabcolsep}{4.0pt}
\tablecaption{Featured Targets\label{tab:obj}}
\tabletypesize{\footnotesize}
\tablehead{
\colhead{Object}                      &
\colhead{$\alpha_{2000}$}             &
\colhead{$\delta_{2000}$}             &
\colhead{Redshift}                    &
\colhead{$N_{\rm spec}$}              &
\colhead{Cadence}                                     
}
\startdata
3C~120       & 04 33 11.1 & $+$05 21 16 & 0.0330 & 27 & 3.6  \\
Ark~120      & 05 16 11.4 & $-$00 08 59 & 0.0327 & 22 & 4.3  \\
Mrk~6        & 06 52 12.2 & $+$74 25 37 & 0.0188 & 45 & 3.4  \\
SBS~1518+593 & 15 19 21.6 & $+$59 08 24 & 0.0781 & 20 & 6.8 \\
\enddata
\tablecomments{$N_{\rm spec}$ is the number of spectroscopic epochs. 
``Cadence'' is the average sampling interval of the objects in days.
}
\end{deluxetable}

\begin{figure*}
\centering
\includegraphics[width=\textwidth]{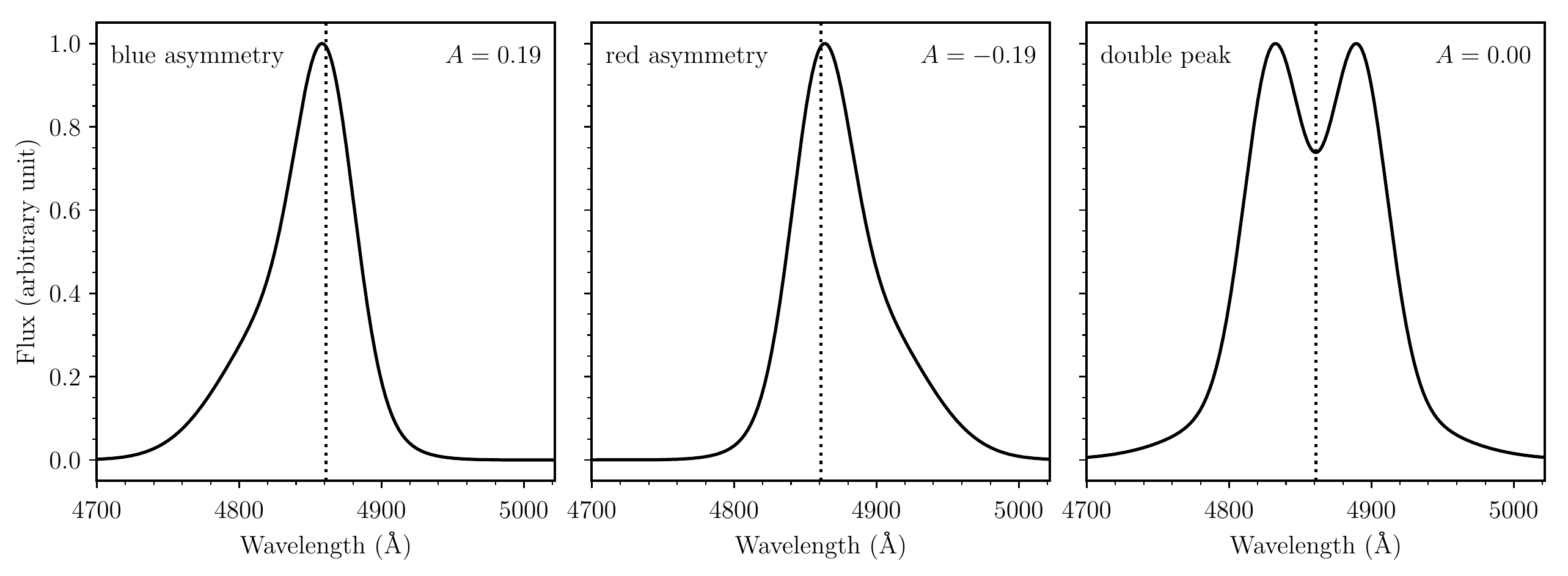}
\caption{Schematic diagram for asymmetric and double-peaked profiles. $A$ is defined as Equation \ref{eqn:asymmetry}. The dotted lines mark 4861\AA. }
\label{fig:asymmetry}
\end{figure*}

Another intriguing hypothesis to explain some asymmetric and shifted profiles is the existence of close binary supermassive black holes (CB-SMBHs) in AGNs.  CB-SMBHs have been predicted to be located in galactic centers due to galaxy mergers \citep{Begelman1980}, however, one single black hole is commonly assumed in explanations of RM data. 
This assumption may be challenged by the more unusual profiles of BELs implying the potential presence of CB-SMBHs
\citep[e.g.,][]{gaskell1996, eracleous1997, eracleous2012, boroson2009, bon2012, bon2016, li2016, li2017, 
decarli2013, shen2013, WangLL2017, runnoe2017, Nguyen2016, Pflueger2018},
which could be an indicator of CB-SMBHs \citep{Popovic2000,Shen2010}. 
However, there are several alternative explanations for the complicated profiles, such as 
elliptical BLR disks for asymmetric or double-peaked profiles \citep{eracleous1997}, 
hot spots \citep{jovanovic2010}, partially covering dusty obscurers \citep{gaskell2018}, 
and even spiral-arms \citep{Storchi-Bergmann2017}. 
Recoiling AGNs may be an additional possibility if the velocity is high enough to
create an appreciable shift from a narrow-line region reflecting the redshift of the host galaxy
\citep{volonteri2008}.

CB-SMBHs are expected to be sources of nano-Hz gravitational waves, likely to be discovered by Pulsar Timing Arrays
\citep[e.g.,][]{sesana2008}.
While it is likely impossible to convincingly diagnose the presence of CB-SMBHs using single epoch spectra, time-series spectroscopy of BELs may prove successful.
Recently, \cite{Wang2018} suggested a high-fidelity RM campaign to determine 
velocity delay maps for AGNs that are binary candidates in order to provide the optical identification of CB-SMBHs.  

For these reasons, we have undertaken the ``Monitoring AGNs with H$\beta$ Asymmetry" (MAHA) project. 
We describe our target selection, observations, and
data reduction in Section \ref{sec:maha}.  We report the initial results for 
four AGNs here, specifically: the mean and rms
spectra in Section \ref{sec:mean_rms}; the measurement of the
light curves in Section \ref{sec:light_curves}; the lag
measurements,  the widths of the \hb\ lines, black hole masses, and the
velocity-resolved time lags in  Section \ref{sec:analysis}. 
We discuss the results for each individual object in Section \ref{sec:discussion}.
Finally, we summarize our findings in Section \ref{sec:summary}.


\begin{deluxetable*}{lccccccccc}
\tablecolumns{10}
\tablecaption{Measurement windows in the observed-frame and \oiii\ standard fluxes\label{tab:windows}}
\tabletypesize{\footnotesize}
\tablehead{
\colhead{}                            &
\colhead{}                            &
\colhead{}                            &
\multicolumn{3}{c}{\oiii}             &
\colhead{}                            &
\multicolumn{3}{c}{\hb}               \\ \cline{4-6} \cline{8-10}
\colhead{Object}                      &
\colhead{$F_{\oiii}$}                 &
\colhead{}                            &
\colhead{continuum (blue)}            &
\colhead{line}                        &
\colhead{continuum (red)}             & 
\colhead{}                            &  
\colhead{continuum (blue)}            &
\colhead{line}                        &
\colhead{continuum (red)}             \\ 
\colhead{}                            &
\colhead{($10^{-13}\ \mathrm{erg\ s^{-1}\ cm^{-2}}$)} &
\colhead{}                            &
\colhead{(\AA)}                       &
\colhead{(\AA)}                       &
\colhead{(\AA)}                       &
\colhead{}                            &
\colhead{(\AA)}                       &
\colhead{(\AA)}                       &
\colhead{(\AA)}                            
}
\startdata
3C~120       & 3.07 & & 5140--5150 & 5150--5200 & 5200--5215 & & 4943--4965 & 4965--5060 & 5088--5103  \\
Ark~120      & 0.84 & & 5140--5150 & 5150--5195 & 5195--5210 & & 4927--4950 & 4950--5138 & 5196--5215  \\
Mrk~6        & 6.32 & & 5070--5080 & 5080--5130 & 5130--5150 & & 4820--4850 & 4850--5076 & 5138--5168  \\
SBS~1518+593 & 0.44 & & 5365--5375 & 5375--5420 & 5420--5435 & & 5175--5196 & 5200--5300 & 5315--5331  
\enddata
\end{deluxetable*}

\section{MAHA: Targets, Observations, Data Reduction}
\label{sec:maha}

Below we describe our MAHA target selection as well as  our program of observations and data reduction that we started in December 2016.  Due to a lack of significant variability, long time lags comparable to the length of an individual seasonal campaign, or other issues, many objects will require multiple campaigns to produce high-quality results of the type we seek.   We expect to add or drop objects as MAHA progresses for a variety of reasons, so the sample should not be considered as absolute.  The sample should, however, illustrate the type of objects we are monitoring and the diversity of their H$\beta$ profiles.  
Our observational methods and data reduction apply generally to the entire sample,
although we provide measurements and analyses in this first paper for only four objects
for which we have obtained good quality velocity-resolved time lags.

\subsection{Targets}

The core MAHA sample includes AGNs with 
asymmetric or double-peaked \hb\ emission lines, as well as objects reported as
 binary black hole candidates (which usually also have asymmetric lines).  
 While the majority of the \hb\ profiles of MAHA targets can easily be visually identified as asymmetric, it will be useful to parameterize asymmetry.
 There exist many ways to quantify asymmetry, each with its pros and cons.
 For ease of historical comparison to previous work \citep[e.g.,][]{DeRobertis1985, boroson1992, brotherton1996}, we adopt the dimensionless asymmetry parameter:
 \begin{equation}
 \label{eqn:asymmetry}
 A = \frac{\lambda_c(3/4) - \lambda_c(1/4) }{\Delta \lambda (1/2)}
 \end{equation}
 where $\lambda_c(3/4)$ and $\lambda_c(1/4)$ are the central wavelengths at the 3/4 and 1/4 of the peak height, and $\Delta \lambda (1/2)$ is the FWHM.
Blue asymmetries are positive, red are negative.  Note that the asymmetry is independent
of any peak or centroid wavelength shift relative to systemic.
 Figure \ref{fig:asymmetry} illustrates what we mean when we say red and blue asymmetries and double-peaked lines.
 
 We selected our targets
 from a variety of literature sources based on both asymmetry measurements and visual
 inspection of optical spectra \citep[e.g.,][]{DeRobertis1985,stirpe1990, boroson1992, eracleous1993,
eracleous1994, marziani2003, hu2008a, hu2008b, eracleous2012}.
We selected some additional sources directly from the quasar sample of the Sloan Digital Sky Survey (SDSS) \citep{schneider2010}, decomposing
the \hb\ lines by two Gaussians through the multi-component fitting procedure
in \cite{hu2008a, hu2008b}, then selecting objects for which the two
Gaussians have substantially different central velocities.  This method recovered many
objects already identified from the literature sources above.

Using the Wyoming
Infrared Observatory (WIRO) 2.3 meter telescope and its longslit spectrograph requires additional
selection criteia.  WIRO has a latitude of 41$^{\circ}$ N, and we do not include targets south of
declination $\sim -10^{\circ}$. In order to obtain sufficiently high signal-to-noise ratios (S/N) in exposure times of $\sim$1 hour or less, the magnitude
(in V or SDSS r$^{\prime}$ band) needs to be brighter than $\sim$17.  
We also require z $< 0.38$ to keep \oiii\ at less than 7000 \AA\ observed frame to 
avoid the inefficiencies of grating tilts and extra calibrations.
Finally, because of our procedure for photometric calibration using narrow lines,
 \oiii $\lambda$5007 cannot be too weak.  
 When everything else is equal, we prefer brighter targets at lower redshifts and more
 northern declinations.

 We do not automatically exclude AGNs from MAHA just because they have previous
 RM results.  BLRs and their corresponding H$\beta$ asymmetries
  may evolve over time periods of several years.  Additionally, we aim to obtain high enough data quality to not only determine velocity-resolved time lags, but also to conduct more involved analyses such as creating velocity-delay maps and dynamical models \citep[e.g.,][]{horne1994, horne2004, bentz2010,
grier2013, skielboe2015, pancoast2011, pancoast2012, pancoast2014b, grier2017vdm}.

 We provide additional information about the MAHA targets and their spectra in Appendix \ref{sec:maha_targets}, including
 the initial MAHA sample, its characteristics, and example WIRO spectra of the H$\beta$ line region obtained during 2016-2018.
 Because of the need to use \oiii\ to calibrate fluxes in WIRO spectra, and the fact that 
 objects with strong \oiii\ tend to be the ones with the strongest red asymmetric \hb\ lines
 \citep{boroson1992}, our sample is biased against objects with blue asymmetric \hb\ lines.
 We plan to search for and add more AGNs with blue asymmetric \hb\ lines as MAHA
 progresses.

As the first paper of the series, we
provide here our RM measurements of 4 AGNs: 3C~120, Ark~120, Mrk~6, 
and SBS~1518+593 (see Table \ref{tab:obj} for their coordinates and some general information).  Their luminosities place them among Seyfert galaxies and broad line 
radio galaxies rather than the more energetic quasar category.  Ark~120 and Mrk~6 show extreme red asymmetric profiles, while 
3C~120 and SBS~1518+593 show milder red asymmetries.  None are 
extreme super-Eddington accretors, although 3C~120 and SBS~1518+593
have dimensionless accretion rates $\dotm$ on order unity
\citep[][see also \ref{sec:BHmass}]{du2015}.
The light curves of all these sources have
shown unambiguous peaks or dips, and are sufficient for us to measure \hb\
time lags.

\subsection{Observations}

\subsubsection{Spectroscopy}

We obtained spectroscopic data using the 2.3 m telescope at the 
WIRO and its Long Slit Spectrograph, observing remotely from the University of Wyoming campus \citep{findlay2016}.
We used the 900 line
mm$^{-1}$ grating, which provides a dispersion of 1.49 \AA\ pixel$^{-1}$ and a
wavelength coverage of 4000 -- 7000 \AA. We adopted a slit width of
5$^{\prime\prime}$ in order to minimize
the light losses due to the aperture effect. We reduced the spectra with
IRAF v2.16, and extracted them using a uniform aperture of $\pm6^{\prime\prime}.8$
and background windows of $7^{\prime\prime}.6$ -- $15^{\prime\prime}.1$ on both
sides. The wavelengths of the spectra are calibrated by taking CuAr
lamp exposures. For each object, we took several consecutive exposures in
order to estimate the systematic flux calibration uncertainties (see Section
\ref{sec:calibration}).

\subsubsection{Spectrophotometry}
\label{sec:calibration}

We initially flux calibrated the spectra using one or more spectrophotometric standard stars
observed each night (primarily Feige 34, G191B2B, or BD+28d4211). 
However, due to variable
atmospheric extinction during the night, we took additional measures to obtain accurate photometry. We used established techniques to use  \oiii\ lines  for flux calibration \citep{vanGroningen1992, fausnaugh2017a}, which ensures good accuracy even in relatively poor observing conditions. The variable time
scales of the \oiii\ $\lambda$5007 lines are much longer than one year for luminous AGNs
as they originate from much more extended narrow-line regions \citep{peterson2013}. 
Therefore, \oiii\ lines can be used as flux standards to calibrate the spectra.

We used a 5$^{\prime\prime}$-wide slit, which is wider than the FWHM of the
seeing ($2^{\prime\prime}\sim3^{\prime\prime}$) during all of the observations. The variable seeing at different times
leads to the change of the line spread function (spectral resolution, see
more details in \citealt{du2016VI}). Additionally, the tracking inaccuracy of
the telescope sometimes made the line spread function deviate
from a Gaussian. Thus, before scaling the spectra according to
their \oiii\ fluxes, we convolved the \oiii\ profiles of the spectra by a
double-Gaussian (a sum of two Gaussians) function ($\zeta$)\footnote{$\zeta$ is not the line
spread function. It represents the differences between the spectral
broadening of the spectrum and the broadest spectral broadening function of
the object.} that fit the broadest \oiii\ profile for each object
 (a Gaussian with the same
FWHM as the broadest \oiii\ was used instead if the broadest \oiii\ itself
significantly differed from Gaussian). We extracted the \oiii\
profiles by subtracting the local continuum underneath
determined by interpolation between two nearby background windows. 
We provide the extraction and local continuum
windows in Table \ref{tab:windows}. The optimal parameters of
$\zeta$ were determined by the Levenberg-Marquardt algorithm. Then, all of the
spectra were smoothed by convolution with their corresponding $\zeta$ to
minimize the influence due to the variable seeing and the tracking inaccuracies. It should be noted that the spectral resolution after the convolution is 
lower than that of the original spectrum, but still quite sufficient to resolve broad \hb\ profiles.

After that, each exposure of the object was scaled to match its standard \oiii\
flux. The \oiii\ fluxes were measured using the windows listed in Table
\ref{tab:windows}, and the standard \oiii\ fluxes (listed in Table
\ref{tab:windows}) of the objects are determined by the spectra taken in the
photometric conditions. We produced the final calibrated spectra by averaging the
(appropriately noise-weighted) exposures in the same night for each object.

\begin{figure*}
\centering
\includegraphics[width=\textwidth]{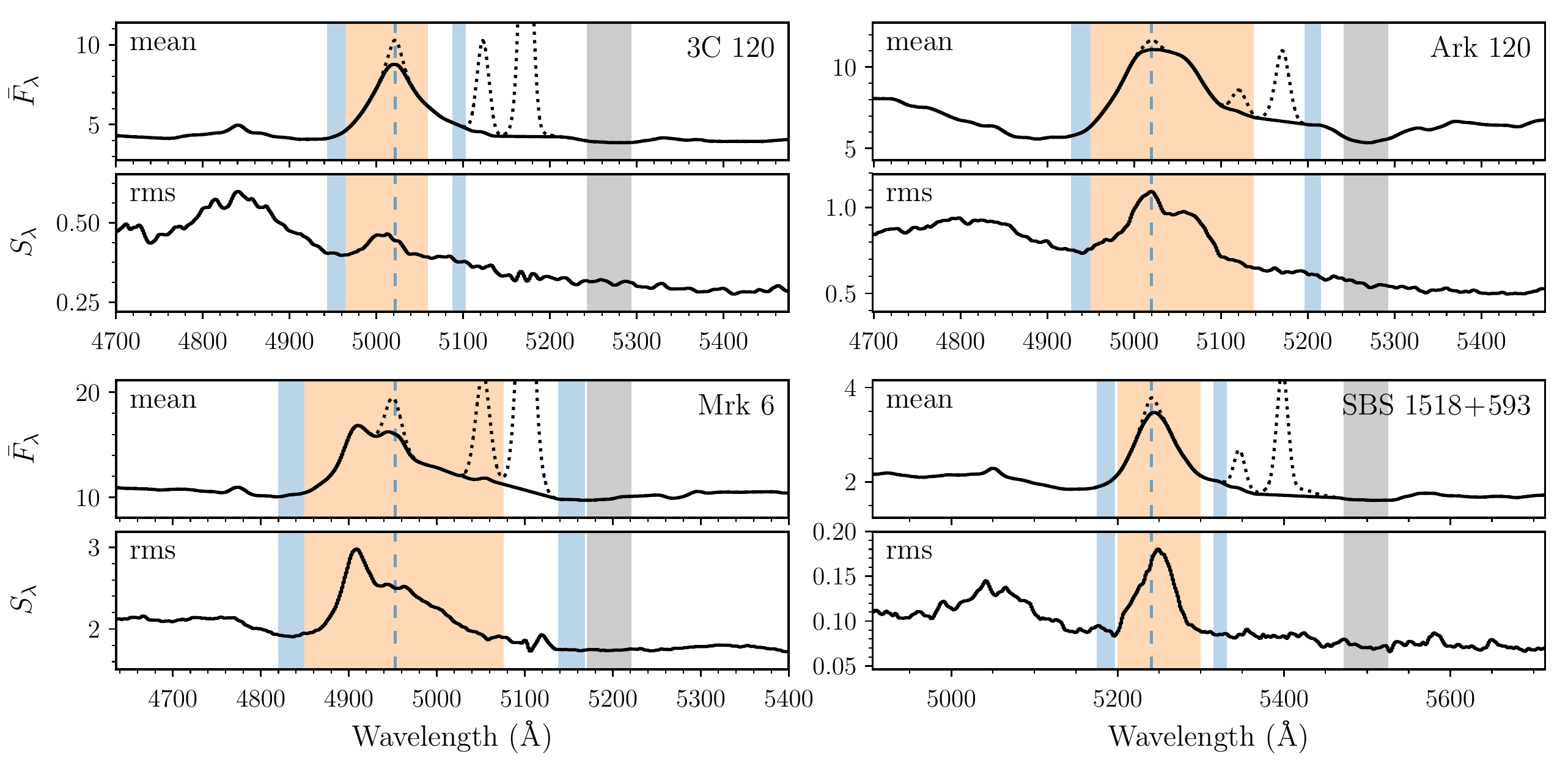}
\caption{Mean and rms spectra in the observed-frame of the objects. The solid lines are the narrow-line-subtracted mean and  
rms spectra, and the dotted lines are the narrow \hb\ and \oiii$\lambda4959,5007$ (see more details in Section \ref{sec:width}). The orange and 
blue regions are the windows for \hb\ emission lines and their backgrounds. The gray regions mark the 5100\AA\ 
continuum windows. We plot $4861\times(1+z)$ \AA\ as blue dashed lines in order to illustrate 
their asymmetric \hb\ profiles more clearly. The flux density units are $10^{-15}~{\rm erg~s^{-1}~cm^{-2}~\AA^{-1}}$.}
\label{fig:meanrms}
\end{figure*}

\subsection{Mean and RMS spectra}
\label{sec:mean_rms}

To demonstrate the spectral characteristics and to evaluate the variation
amplitude at  different wavelengths, we plot the mean and root-mean-square
(rms) spectra for each object in Figure \ref{fig:meanrms}. The mean and rms
spectra are defined as
\begin{equation}
\bar{F}_{\lambda}=\frac{1}{N}\sum_{i=1}^NF_{\lambda}^i
\end{equation}
and 
\begin{equation}
S_{\lambda}=\left[\frac{1}{N}\sum_{i=1}^N\left(F_{\lambda}^i-\bar{F}_{\lambda}\right)^2\right]^{1/2},
\end{equation}
respectively. $F_{\lambda}^i$ is the $i$-th spectrum, and $N$ is the number of spectra for this object. 
The \oiii\ and the narrow \hb\ emission lines in the rms spectra are extremely weak, which indicates that
the calibration procedure in Section \ref{sec:calibration} works very well. 
The obvious broad \hb\ lines in the rms spectra imply that their variations are significant.

\section{Light Curves}
\label{sec:light_curves}

\subsection{Light curves from WIRO}

The fluxes of the \hb\ emission line can be measured by direct
integration \citep[e.g.,][]{peterson1998, kaspi2000, bentz2009,
grier2012, du2014, fausnaugh2017} or by spectral fitting
\citep[e.g.,][]{barth2013, hu2015}. The first method measures  the flux by
integrating the \hb\ line after subtracting the local background determined by
two continuum windows. The second method separates the emission lines from
the continuum by multi-component spectral fitting, and has been gradually
adopted in recent years. Considering that (1) it is difficult to fit the very complex and asymmetric 
\hb\ profiles of our targets perfectly by multiple Gaussian or Lorentzian functions, or their combinations;
(2) the integration method is more robust and works well for isolated emission line like \hb, 
we choose to use the integration method to measure the \hb\ light curves in this work.

We chose the windows for \hb\ flux measurements that cover the \hb\
emission shown in their rms spectra (Figure \ref{fig:meanrms}) and avoid the
possible influences from their \oiii\ lines. The local continuum windows were 
selected as minimally variable regions in the rms spectra. 
We provide the line and the local continuum
windows of \hb\  in Table \ref{tab:windows} and show them in Figure
\ref{fig:meanrms}. We measured the \hb\ light curves by integrating the
fluxes in the \hb\ windows after subtracting the local continuum determined by
interpolating between the two nearby continuum windows. Similar to the narrow \oiii\ lines, the flux of the narrow \hb\ can
be regarded as a constant during our campaign. Thus, we did not remove the contributions of the narrow
 \hb\ lines from the measured \hb\ light curves.
 We obtained the 5100\AA\ continuum light curves by averaging the fluxes from 5075 to 5125\AA\ in the
rest-frames (shown as grey regions in Figure \ref{fig:meanrms}). 

The error bars of the fluxes in the light curves include two components: (1)
the Possion noise and (2) the systematic uncertainties. The systematic
uncertainties result primarily from the variable atmospheric extinction
(especially in poor weather conditions) and telescope-tracking inaccuracies,
and we  estimated them using the scatter of the mean fluxes of the exposures
in the same night over a wider range  of wavelength ($\sim$4700--5100\AA). The
two components summed in quadrature provide  the error bars of the points in
the light curves (see Figure \ref{fig:light_curves}). However, the above error
bars likely do not account for all of the systematic uncertainties for most of
the objects (the flux differences between adjacent nights are sometimes larger
than the error bars and unlikely to result solely from real variability).  We
show additional systematic uncertainties estimated by the median filter
method (see more details in \citealt{du2014}), which smooths the light curve by a
median filter of 5 points and then obtains the systematic uncertainty from the standard deviation 
of the residuals after subtracting the smoothed light curve, if necessary, as the gray
error bars in the lower-right corners in Figure \ref{fig:light_curves}\footnote{The systematic
uncertainties obtained by the median filter (the gray error bars in the corners of Figure \ref{fig:light_curves})
are 0.10 and 0.05, 0.07 and 0.13, and 0.05 and 0.04
for the continuum and emission-line light curves of 3C~120, Ark~120, and Mrk~6, respectively.
The units are $10^{-15}~{\rm erg~s^{-1}~cm^{-2}~\AA^{-1}}$ and $10^{-13}~{\rm erg~s^{-1}~cm^{-2}}$ for
the continuum and emission-line fluxes. For SBS~1518+593, the extra systematic uncertainties are
not necessary. The original error bars are already large enough.}.  
These
are also taken into account in the following time-series analysis (Section
\ref{sec:ccf}) by added in quadrature to 
the error bar of every data point in Figure \ref{fig:light_curves}. We provide the 5100 \AA\ and \hb\ light curves in Figure
\ref{fig:light_curves} and Table \ref{tab:light_curves}.

\begin{deluxetable}{ccccccc}
\tablecolumns{7}
\setlength{\tabcolsep}{3pt}
\tablecaption{Light curves\label{tab:light_curves}}
\tabletypesize{\footnotesize}
\tablehead{
\multicolumn{3}{c}{3C~120}        &
\colhead{}                       &
\multicolumn{3}{c}{Ark~120}      \\ \cline{1-3} \cline{5-7} 
\colhead{JD}                     &
\colhead{$F_{5100}$}             &
\colhead{$F_{\rm H\beta}$}       &
\colhead{}                       &  
\colhead{JD}                     &
\colhead{$F_{5100}$}             &
\colhead{$F_{\rm H\beta}$}                      
}
\startdata
   52.65 & $  4.00\pm  0.02$ & $  2.48\pm  0.01$ & &    43.71 & $  5.28\pm  0.09$ & $  5.66\pm  0.09$\\ 
   71.64 & $  3.98\pm  0.01$ & $  2.55\pm  0.01$ & &    52.68 & $  5.41\pm  0.03$ & $  5.39\pm  0.02$\\ 
   72.59 & $  3.98\pm  0.02$ & $  2.49\pm  0.02$ & &    53.68 & $  5.52\pm  0.09$ & $  5.48\pm  0.09$\\ 
   74.70 & $  4.06\pm  0.01$ & $  2.53\pm  0.01$ & &    71.65 & $  6.07\pm  0.09$ & $  5.54\pm  0.08$\\ 
   80.58 & $  4.14\pm  0.02$ & $  2.51\pm  0.01$ & &    72.61 & $  6.26\pm  0.14$ & $  5.59\pm  0.12$\\ 
\enddata
  \tablecomments{JD: Julian dates from 2,457,700;
   $F_{5100}$ and $F_{\hb}$ are the continuum fluxes at 5100 \AA\ and
     \hb\ fluxes in units of $10^{-15}\ {\rm erg\ s^{-1}\ cm^{-2}\ \AA^{-1}}$ and
      $10^{-13}\ {\rm erg\ s^{-1}\ cm^{-2}}$, respectively. This table is available in its entirety in a 
      machine-readable form in the online journal. A portion is shown here for guidance regarding its form and content.}
\end{deluxetable}

\subsection{Photometric light curves from ASAS-SN}

Photometric observations based on imaging are commonly carried out simultaneously with the
spectroscopic observations in many RM campaigns \citep[e.g.,][]{bentz2009,
denney2010, du2014, du2015, du2016V, du2018, wang2014, fausnaugh2017}. The
photometric light curves can be used to check the calibration precision of the
spectroscopic observations, and furthermore can be adopted as supplemental
to the 5100\AA\ light curves, especially if the sampling of the spectroscopic
observations is relatively poor or their monitoring period is not long enough.
We did not conduct our own photometric observations during 2016--2017 in our
campaign. However, because the objects of this paper are bright enough, we can
find their photometric light curves from archival data from the All-Sky Automated
Survey for Supernovae  (ASAS-SN)\footnote{\url{http://www.astronomy.ohio-
state.edu/~assassin/index.shtml}}. ASAS-SN is a long-term project useful to
discover and study supernovae, transients, and other variable sources by automatic and regular sky survey, and provides photometric light curves for the
objects down to $\sim$17 magnitude. More information and technical details about the 
ASAS-SN light curves are provided by, e.g., \cite{shappee2014} and \cite{kochanek2017}. 
Figure \ref{fig:light_curves} shows the scaled ASAS-SN light curves of our targets (more details of the scaling are provided in
Section \ref{sec:combine}). We removed several points with very poor S/N. 
The variations of the 5100\AA\ and the ASAS-SN light curves are consistent (see Figure \ref{fig:light_curves}), thus it verifies our spectroscopic calibration.

\subsection{Combined continuum light curves}
\label{sec:combine}

Considering that the ASAS-SN observations can extend the temporal spans of our continuum light
curves and improve their sampling cadences, we averaged the ASAS-SN and the
5100\AA\ light curves to produce a combined continuum light curve for each
object. Because of differing apertures (ASAS-SN uses an aperture with a radius of
15$^{\prime\prime}$.6), the ASAS-SN light curves require adjustment to match the
mean fluxes and the variation amplitudes of the WIRO 5100\AA\ light curves 
before combination. This was performed by assuming 
\begin{equation}
\label{eqn:scale}
F_{5100} = a + b \times F_{\rm ASAS-SN}, 
\end{equation}
where $F_{5100}$ and $F_{\rm ASAS-SN}$ are the 5100\AA\ and
ASAS-SN fluxes of the closely adjacent pairs of the observations (the separation
is at most less than 2 days), $a$ is a flux adjustment, and $b$ is a scale factor.
We determined the values of $a$ and $b$ by using the FITEXY algorithm
\citep{press1992}. Then, we scaled the ASAS-SN light curves by applying Equation (\ref{eqn:scale}), 
and combined the 5100\AA\ and ASAS-SN light curves by weighted averaging all of the observations in the same nights.
The uncertainties of the weighted mean\footnote{The uncertainties of the weighted mean is 
defined as $\sigma_{\rm mean}=(1/\sum\sigma_i^{-2})^{1/2}$, where $\sigma_i$ is the uncertainty of each point in the same night.} are simply adopted as the error bars in the combined light curves. 
We tried to use the median filter method to estimate the systematic uncertainties, and found no extra systematic uncertainties 
are needed for the combined light curves.
Figure \ref{fig:light_curves} shows both the scaled ASAS-SN light curves and the final combined light curves.


\begin{deluxetable}{lcclclc}
  \tablecolumns{6}
  \tablecaption{Time lags\label{tab:lags}}
  \tabletypesize{\scriptsize}
  \tablehead{
      \colhead{}                         &
      \colhead{}                         &
      \colhead{}                         &
      \colhead{Observed}                 &
      \colhead{}                         &
      \colhead{Rest-frame}               &
      \colhead{}                         \\
      \colhead{Object}                   &
      \colhead{Continuum}                &
      \colhead{$r_{\rm max}$}            &
      \colhead{Time Lag}                 &
      \colhead{}                         &
      \colhead{Time Lag}                 &
      \colhead{Note}                     \\ 
      \colhead{}                         &
      \colhead{}                         &
      \colhead{}                         &
      \colhead{(days)}                   &
      \colhead{}                         &
      \colhead{(days)}                   &
      \colhead{}                         
            }
\startdata
3C~120       & 5100\AA  & 0.76 & $20.9_{- 4.4}^{+ 5.1}$ & & $20.2_{- 4.2}^{+ 5.0}$ & $\surd$ \\
             & combined & 0.71 & $19.3_{- 3.0}^{+ 3.0}$ & & $18.7_{- 2.9}^{+ 2.9}$ &         \\
Ark~120      & 5100\AA  & 0.94 & $16.7_{- 3.2}^{+ 3.3}$ & & $16.2_{- 3.1}^{+ 3.2}$ & $\surd$ \\
             & combined & 0.83 & $13.8_{- 2.8}^{+ 3.1}$ & & $13.4_{- 2.7}^{+ 3.0}$ &         \\
Mrk~6        & 5100\AA  & 0.99 & $40.4_{-11.4}^{+ 5.2}$ & & $39.6_{-11.2}^{+ 5.1}$ &         \\
             & combined & 0.99 & $18.8_{- 2.5}^{+ 2.5}$ & & $18.5_{- 2.4}^{+ 2.5}$ & $\surd$ \\
SBS~1518+593 & 5100\AA  & 0.93 & $21.2_{- 6.4}^{+10.7}$ & & $19.7_{- 6.0}^{+ 9.9}$ & $\surd$ \\
             & combined & 0.88 & $30.6_{- 7.3}^{+ 8.9}$ & & $28.4_{- 6.8}^{+ 8.3}$ &         
 \enddata
  \tablecomments{\footnotesize
  ``$\surd$'' means we use this time lag of the object to calculate its black hole mass in Table \ref{tab:fwhm_mbh}.}
\end{deluxetable}

\begin{figure*}
\centering
\includegraphics[width=0.75\textwidth]{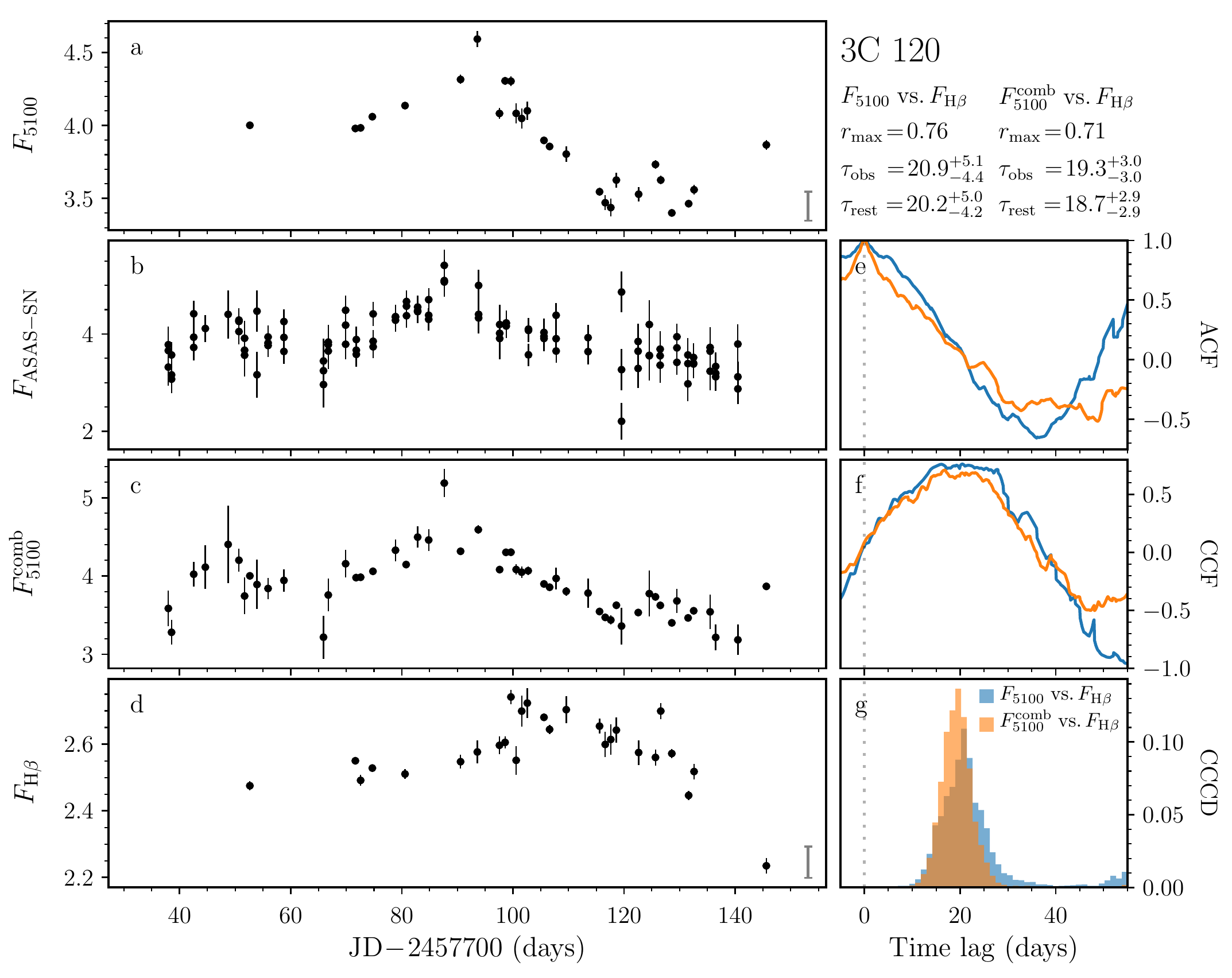} \\
\includegraphics[width=0.75\textwidth]{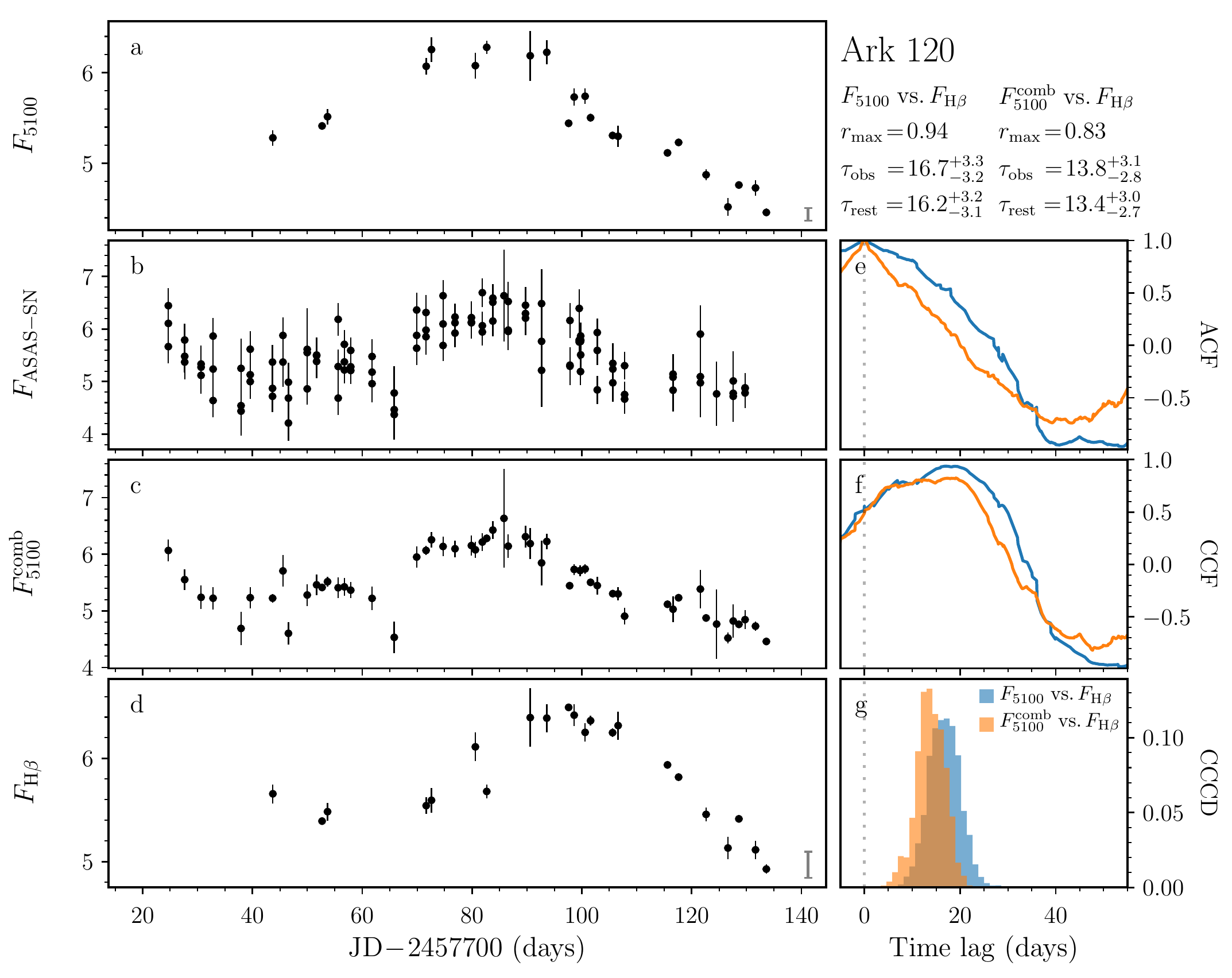} 
\caption{Light curves and cross-correlation functions. Panel a, b, c, and d
are the continuum at 5100\AA, photometry from ASAS-SN,  combined continuum,
and \hb\ light curves. Panel e, f, and g are ACF, CCF, and CCCD. The name of
the object and the corresponding  lag measurements in both of the observed and
rest frames are marked in the upper-right corner of the figure. The units of
the continuum and emission-line light curves are ${10^{-15}\ \rm erg\ s^{-1}\
cm^{-2}\ \AA^{-1}}$ and ${10^{-13}\ \rm erg\ s^{-1}\ cm^{-2}}$, respectively.
Systematic errors are added if needed and shown as the gray error bars in the
lower-right corners in panels a -- d (see more details in Section
\ref{sec:light_curves}).  } 
\label{fig:light_curves} 
\end{figure*}

\begin{figure*}
\figurenum{\ref{fig:light_curves}}
\centering
\includegraphics[width=0.75\textwidth]{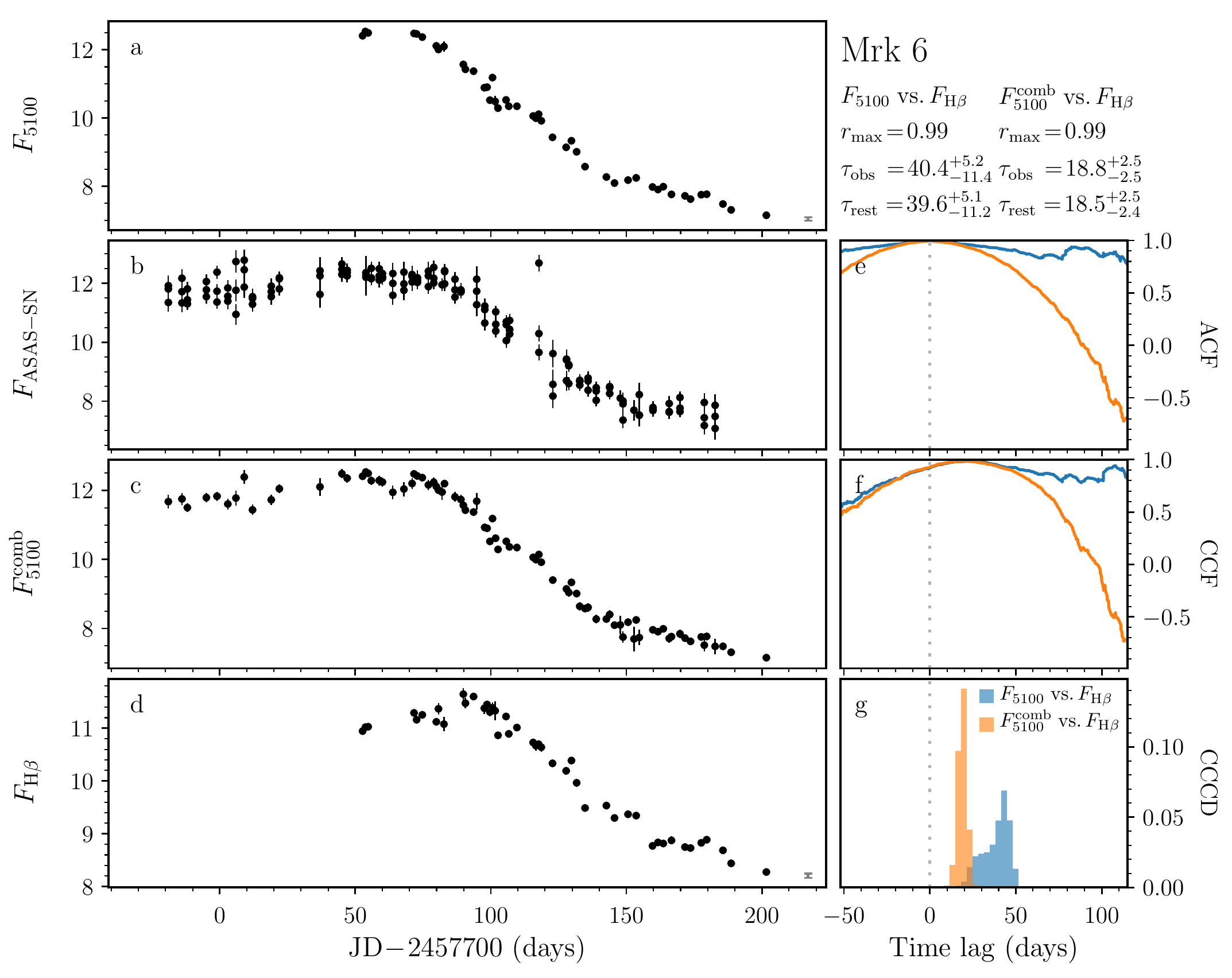} \\
\includegraphics[width=0.75\textwidth]{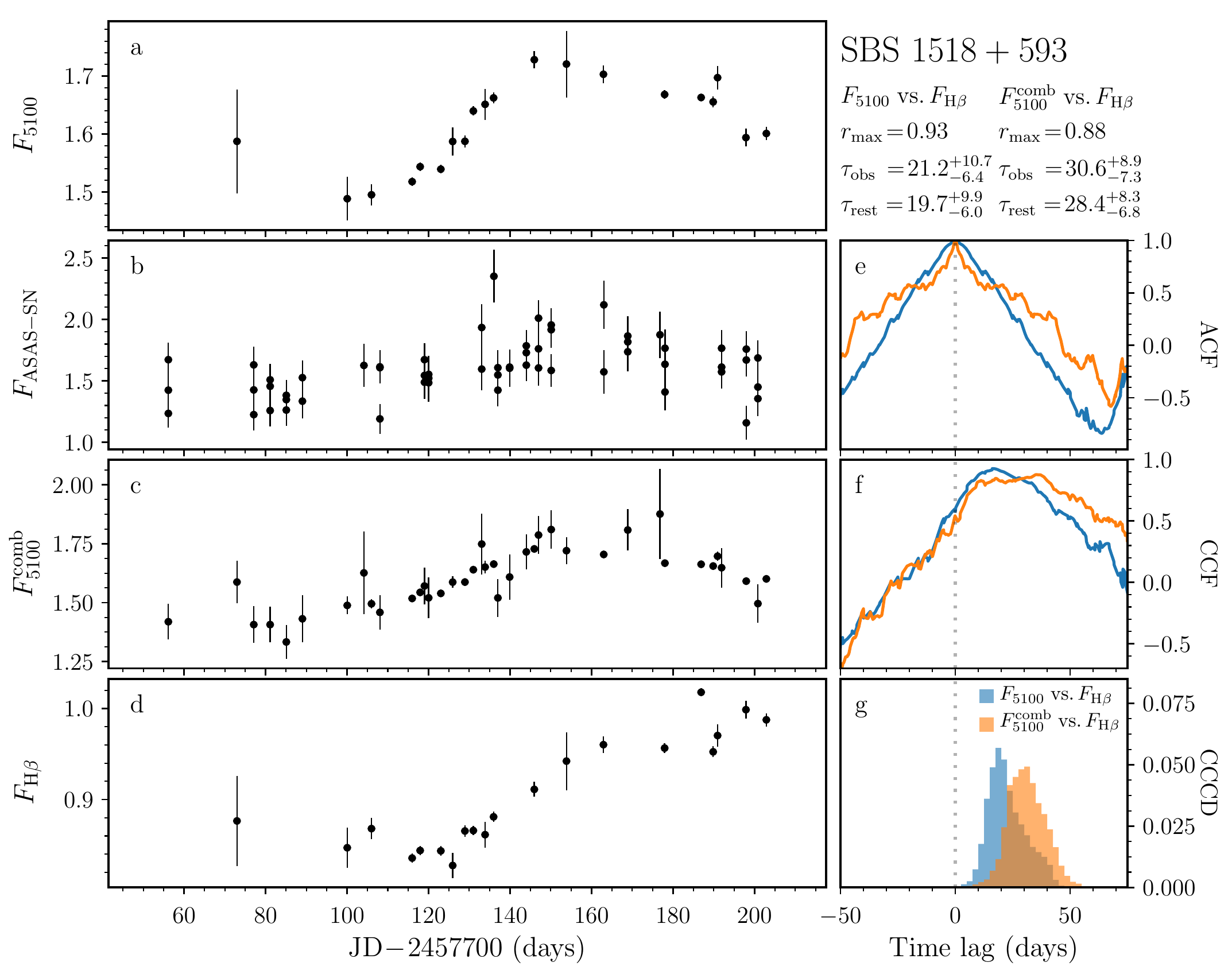} 
\caption{(Continued.)}
\end{figure*}

\section{Analysis}
\label{sec:analysis}

\subsection{Cross-correlation function}
\label{sec:ccf}

The time delays between the variations of the continuum and \hb\ emission
lines were determined by using the interpolated cross-correlation function
(ICCF; \citealt{gaskell1986, gaskell1987}). To measure the time delay we 
used the centroid of the ICCF
above 80\% of the peak (a typical value used in many RM investigations, e.g.,
\citealt{bentz2009, du2014, fausnaugh2017}).

We estimated the uncertainties of the time delays through the
``flux randomization/random subset sampling (FR/RSS)'' method
\citep{peterson1998, peterson2004}. The procedure takes into account both the measurement
errors of the fluxes and the uncertainties due to the sampling/cadence. This
method generates light curve realizations by perturbing the fluxes in
accordance with their error bars and randomly sub-sampling the data points in
the light curves. The cross-correlation centroid distributions (CCCDs) are
obtained by performing the ICCF to the light curve realizations. The median and
the 68\% confidence intervals of the CCCDs are adopted as the final time lags
and their uncertainties.  
The auto-correlation functions (ACFs), the CCFs, and the CCCDs of the 5100\AA\
and \hb\ light curves for each object are shown in Figure
\ref{fig:light_curves}. 
Table \ref{tab:lags} gives The time lag measurements and the corresponding maximum correlation coefficients of the CCFs. We also measured the \hb\ time delays relative to the
combined continuum light curves (see Section \ref{sec:combine}), and provide
the results in Figure \ref{fig:light_curves} and in Table \ref{tab:lags}. 

Although the photometric data can in principle extend the temporal spans of
the 5100\AA\ light curves and improve the sampling cadences, the scatter of
the ASAS-SN light curves are generally larger than that of the 5100\AA\ light
curves in our campaign because of the limited collecting area of the ASAS-SN telescopes. 
We used the lag measurements of the combined continuum versus the \hb\ light
curves only in the case of Mrk 6, which greatly extended continuum coverage 
and resulted in a clearly improved result. The \hb\ time lags we used in the
following analysis are labeled by ``$\surd$'' in Table \ref{tab:lags}.

\begin{deluxetable}{lcc}
\tablecolumns{3}
\tablewidth{\textwidth}
\setlength{\tabcolsep}{5pt}
\tablecaption{Narrow-line flux ratios\label{tab:narrow_ratios}}
\tabletypesize{\footnotesize}
\tablehead{
\colhead{Object}                     &
\colhead{\oiii$\lambda$5007/\oiii$\lambda$4959}             &
\colhead{\oiii$\lambda$5007/\hb}                        
}
\startdata
3C~120  & 3.06$\pm$0.04 & 11.16$\pm$0.40 \\
Ark~120 & 3.20$\pm$0.07 &  8.65$\pm$0.46 \\
Mrk~6   & 3.15$\pm$0.05 &  9.19$\pm$0.31 \\
SBS~1518+593 & 3.18$\pm$0.04 & 7.74$\pm$0.29
\enddata
\end{deluxetable}

\subsection{Line-width measurements}
\label{sec:width}

We measured the line width (using both the FWHM and line dispersion
$\sigma_{\hb}$ parameters)  of the broad \hb\ for both the mean and rms
spectra.  The narrow-line contributions in the rms spectra are fairly weak for
all of the targets, thus the corresponding FWHM and $\sigma_{\hb}$ of the rms
spectra can be obtained easily.  We measured the \hb\ profiles after subtracting
continuum; continuum windows were selected beyond any contribution from the emission lines and can differ between the mean and rms profiles and from the ones listed in Table \ref{tab:windows}, which were optimized for light curves. 
However, before measuring the \hb\ widths from the mean
spectra, the narrow \hb\ and \oiii$\lambda4959,5007$ lines still do need to be
removed.

We first assumed that all of the three narrow lines (\hb\ and \oiii$\lambda4959,5007$) have the same profile. Since the blue wing of \oiii$\lambda5007$  can be slightly blended with the
red wing of \oiii$\lambda4959$ (see Figure \ref{fig:meanrms}), we started by removing
the \oiii$\lambda4959$ by shifting and scaling the \oiii$\lambda5007$ line
(after subtracting the local continuum background determined from the
interpolation of two nearby windows). Then, the corrected \oiii$\lambda5007$
profile was extracted from the \oiii$\lambda4959$-subtracted spectrum, and was
treated as the template of the narrow-line profile. We obtained the narrow component of
the \hb\ line by fitting the spectrum in a very narrow and local window around 4861\AA\ (not the entire \hb\ profile)
with the narrow-line template, a Gaussian (a second Gaussian was added if
necessary), and a linear background. The Gaussian(s) $+$ 
 linear background is used to account for the contribution from the broad \hb\ in the narrow window (the top part of the broad
 \hb). 
 Because the 
 \hb\ profile in the present sample is very complex and asymmetric, we did not fit the entire profile but 
  a narrower and local window (usually a window of $4000\sim5000$ km~s$^{-1}$) around 4861\AA. We tried to add more Gaussians or changed the Gaussian(s) to a high-order polynomial as
  the broad \hb\ contribution in the fitting, but the fitting
  results and the following width measurements did not change significantly because the shape of the narrow template has been constrained by the corresponding \oiii$\lambda5007$ profile of each object. Table \ref{tab:narrow_ratios} provides the narrow \oiii$\lambda5007$/\hb\ and \oiii$\lambda5007$/\oiii$\lambda4959$ flux ratios obtained by this procedure, and Figure \ref{fig:meanrms} shows the narrow-line-subtracted
\hb\ profiles for the mean spectra. The \oiii$\lambda5007$/\hb\ and
\oiii$\lambda5007$/\oiii$\lambda4959$ ratios are close to the typical values
(3 and 10, respectively, e.g., \citealt{kewley2006, stern2013}) in AGNs, which
means that the narrow-line subtraction procedure adopted here appears to work well.
Then we measured the FWHM and $\sigma_{\hb}$ from the mean \hb\ profiles
after the narrow-line subtraction. 

We estimated the broad \hb\ velocity width 
uncertainties for both the mean spectra and the rms spectra by using the bootstrap method. 
A subset spectrum was created by resampling $N$ points with
replacement from the $N$ data points in the mean/rms spectrum.  For the rms
spectrum, we measured the FWHM and $\sigma_{\hb}$ from the subset spectrum, and
repeated the resampling and the measurement 500 times to generate the FWHM and
$\sigma_{\hb}$  distributions. The resulting median values and the standard
deviations of the distributions were regarded as the measurements and the
uncertainties. For the mean spectrum, we subtracted the narrow lines from the
subset spectrum by using the procedure described above before the
measuring the widths. The uncertainties of the narrow-line flux ratios in 
Table \ref{tab:narrow_ratios} were also obtained at the same time.

To estimate the width of the line spread function (instrumental broadening) in
our observations, we measured the widths of the \oiii\ lines in the mean spectra
and compared them with  the intrinsic narrow-line widths in
\cite{whittle1992} or the higher resolution spectra of the Sloan Digital Sky
Survey (SDSS). The FWHM of the line spread function obtained for different
object ranges from $\sim$850 km~s$^{-1}$ to $\sim$1000 km~s$^{-1}$. 
For simplicity, we adopted
the mean value of 925 km/s (FWHM) as the line spread function for all of
the objects in our campaign. After correcting the contribution of the line
spread function, the line widths (FWHM and $\sigma_{\hb}$) of the \hb\ in the
mean and rms spectra are listed in Table \ref{tab:fwhm_mbh}.


\begin{deluxetable*}{lcccccccccc}
  \tablecolumns{11}
  \tablecaption{\hb\ Width Measurements and Black Hole Masses\label{tab:fwhm_mbh}}
  \tabletypesize{\scriptsize}
  \tablehead{
      \colhead{}                   &
      \multicolumn{2}{c}{mean spectra}   &
      \colhead{}                         &
      \multicolumn{2}{c}{rms spectra}      &  
      \colhead{}                         &
      \colhead{$M_{\rm VP}$ (mean spectra)}                    &
      \colhead{}                 &
      \multicolumn{2}{c}{$M_{\bullet}$ (rms spectra)}                 \\ \cline{2-3} \cline{5-6} \cline{8-8} \cline{10-11}
      \colhead{Object}                         &
      \colhead{$V_{\rm FWHM}$}                     &
      \colhead{$\sigma_{\rm line}$}           &
      \colhead{}                         &
      \colhead{$V_{\rm FWHM}$}                     &
      \colhead{$\sigma_{\rm line}$}           &
      \colhead{}                         &
      \colhead{$R_{\rm H\beta} V_{\rm FWHM}^2/G$}                    &
      \colhead{}                 &
      \colhead{$1.12\!\times\!R_{\rm H\beta} V_{\rm FWHM}^2/G$}                 &
      \colhead{$4.47\!\times\!R_{\rm H\beta} \sigma_{\rm line}^2/G$}                 \\
      \colhead{}                         &
      \colhead{(km s$^{-1}$)}            &
      \colhead{(km s$^{-1}$)}            &
      \colhead{}                         &
      \colhead{(km s$^{-1}$)}            &
      \colhead{(km s$^{-1}$)}            &
      \colhead{}                         &
      \colhead{($10^7 M_{\odot}$)}                         &
      \colhead{}                 &
      \colhead{($10^7 M_{\odot}$)}                         &
      \colhead{($10^7 M_{\odot}$)}                         }
\startdata
       3C~120  & $3711\pm34$ & $3174\pm32$ & &  $2343\pm26$ & $1360\pm42$ & & $ 5.43_{-1.13}^{+1.35}$ & & $ 2.43_{-0.51}^{+0.60}$ & $ 3.26_{-0.71}^{+0.83}$\\
      Ark~120  & $6487\pm16$ & $3929\pm14$ & &  $5247\pm25$ & $2184\pm31$ & & $13.31_{-2.55}^{+2.63}$ & & $ 9.75_{-1.87}^{+1.93}$ & $ 6.75_{-1.30}^{+1.35}$\\
        Mrk~6  & $5457\pm16$ & $3647\pm29$ & &  $5274\pm22$ & $3300\pm30$ & & $10.76_{-1.40}^{+1.46}$ & & $11.25_{-1.46}^{+1.52}$ & $17.59_{-2.30}^{+2.40}$\\
 SBS~1518+593  & $3374\pm17$ & $2429\pm37$ & &  $2499\pm22$ & $1038\pm20$ & & $ 4.38_{-1.33}^{+2.20}$ & & $ 2.69_{-0.82}^{+1.35}$ & $ 1.85_{-0.57}^{+0.93}$
 \enddata
  \tablecomments{The line spread function caused by the instrument and seeing has been corrected from the line-width measurements.
  $M_{\rm VP}$ is the virial product measured from the mean spectrum (see more details in Section \ref{sec:BHmass}). The black hole 
  masses (\mbh) are calculated from the rms spectra using the \fblr\ factors in \cite{woo2015}. 
      }
\end{deluxetable*}

\subsection{ Black hole masses and accretion rates}
\label{sec:BHmass}

Combining the time lag with the line width measured from the FWHM or line dispersion, the black hole mass can be obtained by the
Equation \ref{eqn:mass}. 
For AGNs with extreme BLR kinematics, e.g., super-fast or ultra-strong inflow
or outflow, or extreme inclination angles, the virial factor $\fblr$ may possibly differ significantly from typical values \citep[e.g.,][]{pancoast2014b}.

The
mean $f_{\rm BLR}$ of a sample can be calibrated by comparing the RM objects,
which have bulge stellar velocity dispersion measurements ($\sigma_{*}$), with
the $M_{\bullet}$--$\sigma_{*}$ relation of the inactive galaxies (e.g.,
\citealt{onken2004, woo2010, woo2015, park2012, grier2013f, ho2014}, see a
brief review in \citealt{du2017}). However,  each individual object may have
a very different virial factor \citep[e.g.,][]{horne1994, horne2004, bentz2010,
grier2013,pancoast2014b}, especially for those AGNs with asymmetric \hb\ profiles which may
host BLRs with complex geometry or kinematics. We adopt $f_{\rm
BLR}=1.12,\,4.47$ in \cite{woo2015} corresponding to the FWHM and $\sigma_{\rm
H\beta}$ in  the rms spectra, respectively, and also provide simple virial
products (assuming $f_{\rm BLR}=1$) for the FWHM measurements in the mean
spectra for the present sample in Table \ref{tab:fwhm_mbh}, but acknowledge 
the potentially large uncertainty on $f_{\rm BLR}$. The uncertainties of the black hole masses listed 
in Table \ref{tab:fwhm_mbh} only account for the error bars of the lag and width measurements.

We provide general estimates of their dimensionless accretion rates, defined as $\dotm =
\dot{M}_{\bullet}/L_{\rm Edd} c^{-2}$, where $\dot{M}_{\bullet}$ is the mass
accretion rate and $L_{\rm Edd}$ is the Eddington luminosity \citep{du2015,
du2016V}. The \dotm\ can be estimated by using the formula \citep[see more details in][]{du2015,
du2016V}
\begin{equation}
\dotm = 20.1 \left(\frac{\ell_{44}}{\cos i}\right)^{3/2} m^{-2}_7,
\end{equation}
where $\ell_{44} = L_{5100} / 10^{44}\ {\rm erg\ s^{-1}}$ is the
monochromatic luminosity at 5100 \AA, $m_7 = \mbh / 10^7 M_{\odot}$, and $i$ is the  inclination angle of disk to
the line of sight. We adopted $\cos i = 0.75$ (see some discussions in \citealt{du2016V}), 
which is an average estimate for type I AGNs (e.g., \citealt{pancoast2014b}). For the most
precise results, it is necessary to subtract
the host contribution from $L_{5100}$ before calculating the accretion rates, but this 
is beyond the present scope of this paper, so our estimates are upper limits.
We found $\dotm\lesssim1\sim2$ for 3C~120 and SBS~1518+593, and $\lesssim0.2$ for Ark~120 and Mrk~6. Therefore, all of the 4 objects are sub-Eddington accretors.  More detailed determinations of luminosity will be done in a future paper allowing more precise estimates.

\subsection{Velocity-resolved time lags}
\label{sec:velocity_resolved_lags}

In order to investigate the geometry and kinematics of their BLRs, we measure
the velocity-resolved time lags of the \hb\ emission lines of our target AGNs.  
Several typical transfer function models
and their corresponding velocity-resolved time lags are given by, e.g., \cite{bentz2009}, 
and more complicated examples of the transfer function are provided by, e.g., \cite{welsh1991} and \cite{horne2004}.
In general, longer lags in the high-velocity blue/red part of the emission line are regarded as the signatures of inflow/outflow, while a symmetric velocity-resolved structure around zero velocity, with smaller time lags for higher velocities, is diagnostic of Keplerian disk or at least virialized motion over a spatially extended BLR in general.
We divided the \hb\ lines into several velocity bins, each of which having the same integrated fluxes in their individual continuum-subtraced rms spectra based on interpolation between the
windows shown in Figure \ref{fig:meanrms}. Then, we measured
the light curve in each bin and performed an ICCF analysis (using the method in
Section \ref{sec:ccf}) with the 5100\AA\ continuum light curve (the combined continuum light curve in the case of Mrk~6). Figure \ref{fig:v_resolved_lags}
shows the resulting time lags as a function of the velocity and the corresponding rms spectrum for each object. 

The velocity bins with equal rms flux have the same level of variation
but may have different amounts of physical flux. As a further test, we divided
the \hb\ lines into the velocity bins, each of which having the same \hb\
fluxes in the narrow-line-subtracted mean spectra, and measured the velocity-resolved lags in
Appendix \ref{sec:v_resolved_lag_mean}. In general, the results are almost the
same as the rms-based velocity-resolved lags, which means the velocity-resolved 
analysis here is robust. In the following Section \ref{sec:discussion}, the discussion 
of their BLRs is based on the results in Figure \ref{fig:v_resolved_lags}.

\section{Discussion}
\label{sec:discussion}

Our data have produced integrated \hb\ time lags as well as 
high-quality velocity-resolved time lags for our four featured AGNs.  
We discuss each object individually below and compare our results to past 
work as appropriate.

\subsection{Individual Objects}\label{sec:individual}

\subsubsection{3C~120}

\label{sec:3c120}

\begin{figure*}
\centering
\includegraphics[width=\textwidth]{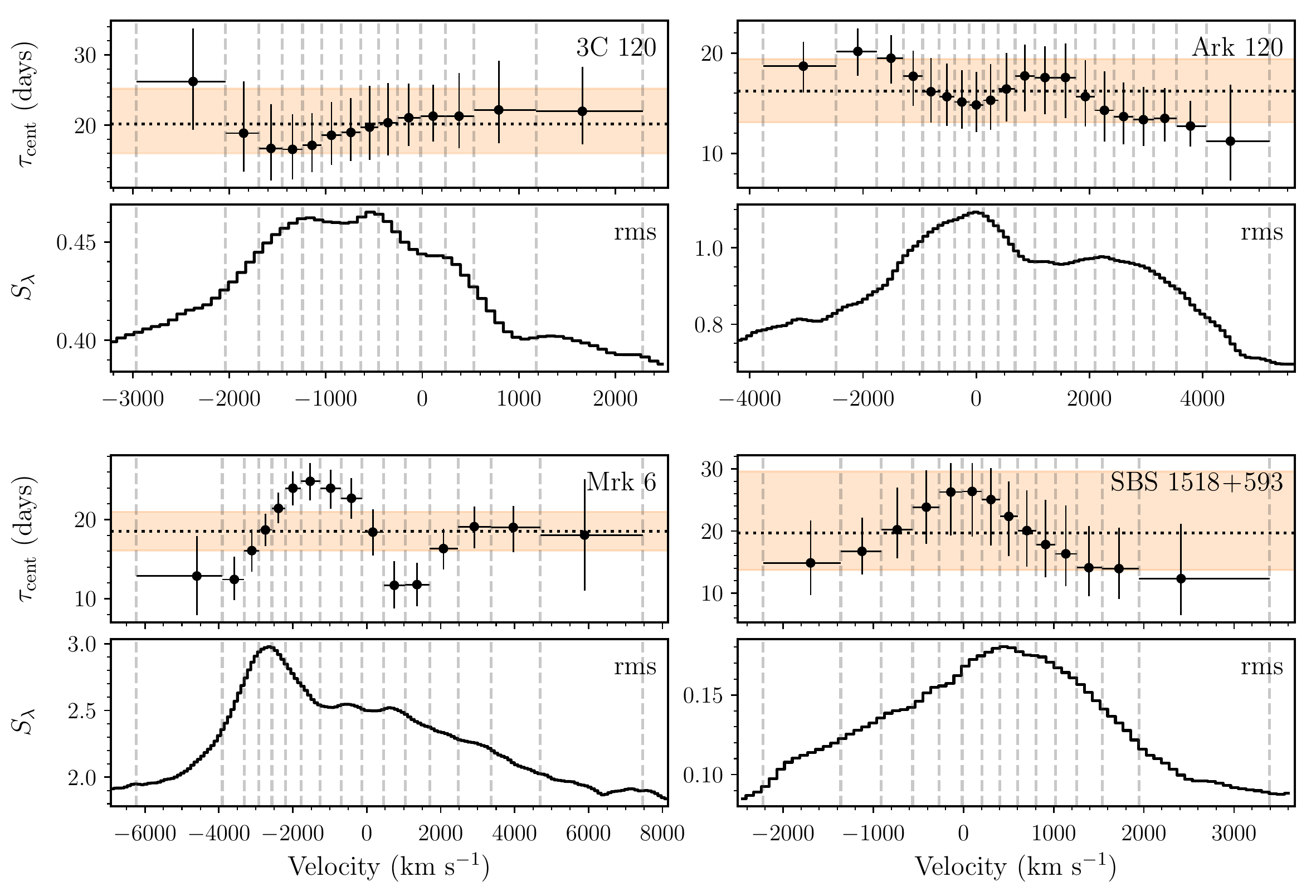}
\caption{Rest-frame velocity-resolved time lags and the corresponding rms spectra. The upper panel in each plot shows
the centroid lags at different velocities, and the lower panel is the rms spectrum in unit of ${10^{-15}\ \rm erg\ s^{-1}\
cm^{-2}\ \AA^{-1}}$. The name of the object is marked in the upper-right corner of each plot. The vertical dashed lines are the 
edges of velocity bins. The horizontal dotted lines and the orange color are the average time lags and the uncertainties 
in Table \ref{tab:lags}. }
\label{fig:v_resolved_lags}
\end{figure*}

As a nearby broad-line radio galaxy, its mean \hb\ profile is
 asymmetric toward the red, however the rms \hb\ profile is
strongly blueshifted  (see Figures \ref{fig:meanrms} and \ref{fig:v_resolved_lags}).  
Its \hb\ time lag
respect to the varying continuum was first detected successfully by
\cite{peterson1998}, albeit with large uncertainty, and the measured lag was $43.8_{-20.3}^{+27.7}$ days in
the rest frame. After $\sim11$ years, it was monitored again in 2008 -- 2009,
and the observed rest-frame \hb\ time lag was $27.9_{-5.9}^{+7.1}$ days
\citep{kollatschny2014}. 3C 120 was observed the third time with higher temporal
sampling in 2010 -- 2011, and a similar \hb\ delay of $25.9_{-2.3}^{+2.3}$ days in
the rest frame was obtained from the light curves by using CCF analysis
method \citep{grier2012}. From the velocity-resolved time lag measurement and
the transfer function reconstructed by the maximum entropy method
\citep{grier2013}, its BLR was likely an inclined disk or a spherical shell,
but there was also some evidence of inflow given that the strength of its line response was
asymmetric \citep{grier2013}.

Our campaign was carried out $\sim$7 years later than the observation in
\cite{grier2012}, and captured a very strong peak around Julian date $\sim$90 days (from the zero
point of 2457700 in Figure \ref{fig:light_curves}) in
the continuum and a clear response around $\sim$110 days in the \hb\ light curve. 
The detected \hb\ lag is $20.2_{-4.2}^{+5.0}$ days in the rest frame, which is slightly 
shorter than the value in \cite{grier2012}, but the difference is not significant considering
the uncertainties. In addition, the rms spectrum in our campaign is different from
that in \cite{grier2012}. Our rms spectrum is significantly blueshifted (see Figure \ref{fig:v_resolved_lags}), 
while the rms spectrum in \cite{grier2012} has a strong red asymmetry. This implies that the 
varying part of the BLR in 3C~120 has significantly changed after $\sim$7 years. 

Furthermore, 
the velocity-resolved lag measurement shows a complicated structure, which is different 
from the symmetric velocity-resolved results of a inclined disk or a spherical shell 
determined by
\cite{grier2013}. From $1500$ km~s$^{-1}$ to $-1500$ km~s$^{-1}$, the lag gradually decreases, 
which is the signature of outflow. However, the tendency changes around $\sim1500$ km~s$^{-1}$, 
and the lags become longer at the blue end. This complicated structure suggests that its BLR 
is in a complex state. Of course, it should be noted that only the two bins (with a little larger uncertainties) 
at the highest blue velocities
are in charge of the upturn at the blue side. More observations with better spectral resolution and higher S/N ratios
are needed in the future to verify this complex state and to investigate the detailed BLR kinematics.

\subsubsection{Ark~120}
\label{sec:ark120}

Given its asymmetric/double-peak \hb\ profile and the long-term periodic-like
variation in  past decades,  Ark~120 is a possible candidate for a
supermassive black hole binary system \citep{li2017}. Its \hb\ time lags
during 1990 September -- 1991 March and 1995 September -- 1996 April, measured
by \cite{peterson1998, peterson2004}, were $47.1^{+8.3}_{-12.4}$ days and
$37.1^{+4.8}_{-5.4}$ days in the rest frame, respectively.
\cite{doroshenko2008} monitored this object from 1992 to 2005, and obtained an 
\hb\ lag of $70\pm7$ days in the rest frame, by combining their data with
other additional data from 1988 to 1996 \citep{peterson1998}. 

During our campaign, the mean spectrum of Ark~120 was not significantly double-peaked, 
but the rms spectrum shows clearly two peaks. One is located at roughly
4861\AA\ in the rest frame,  and the other is strongly redshifted by
$\sim2500$ km~s$^{-1}$. The rms spectrum is different  from the one in
\cite{peterson1998}. The rms spectrum in \cite{peterson1998} has three peaks
(an additional blueshifted peak), and the redshifted peak is relatively stronger. 
One simple and reasonable guess is that both of the blueshifted and redshifted peaks
became relatively weaker with respect to the peak with zero velocity by the epoch of our campaign, which implies a possible correlation between the origins of these two peaks. 

Our velocity-resolved lag measurement of Ark~120 is also complicated. In
general, the time lag decreases from the blue ($-3000$ km~s$^{-1}$) to the red
($4000$ km~s$^{-1}$) velocity, which is the signature of an inflowing BLR.
However,  there is a local peak around $1000\sim2000$ km~s$^{-1}$, which
corresponds to the dip between the two peaks in the rms spectrum. To further
investigate the geometry  and kinematics of the BLR in Ark~120, reconstructing
its velocity-delay map is needed. We will reconstruct the velocity-delay map
of the objects in the present sample using the maximum entropy method
\citep[e.g.,][]{horne1994} in a separate paper in the near future.

\subsubsection{Mrk~6}
\label{sec:mrk6}

It has been known for more than 40 years that the \hb\ profile of Mrk~6
is extremely  asymmetric and has a blueshifted peak with the velocity of
$\sim-3000$ km~s$^{-1}$ \citep{khachikian1971}. With its strange radio
morphology (jet flips and jet precession as reported by, e.g., \citealt{kukula1996,
kharb2014}), Mrk~6 is suggested to be a potential  supermassive binary black hole
system \citep{kharb2014}.

Mrk~6 has been spectroscopically monitored by different groups in the  past
decades \citep{sergeev1999, doroshenko2003, doroshenko2012, grier2012}. With
the long temporal span of the campaign (1992 -- 2008), \cite{doroshenko2012}
found the time lag of its \hb\ line to be $21.1\pm1.9$ days in the rest frame.
\cite{grier2012} observed this object again in 2010 with a higher sampling rate, and
obtained a different \hb\ lag of $10.1\pm1.1$ days. In our observation, the
new \hb\ time lag is $18.5_{-2.4}^{+2.5}$ days in the rest frame,  which is
more similar to the result of \cite{doroshenko2012}. Moreover, the line width
(FWHM$=5274$ km~s$^{-1}$) in our rms spectrum is similar to the value (5445
km~s$^{-1}$) in \cite{doroshenko2012}, and much smaller than the number (9744
km~s$^{-1}$) in  \cite{grier2012}. These changes might be simply due to the 
BLR ``breathing'' effect \citep[e.g.,][]{peterson2002, korista2004, cackett2006}.

Considering the very complex \hb\ profile of this object, it is interesting to
compare the rms spectrum and the velocity-resolved time lags in our campaign
with those reported by \cite{doroshenko2012} and \cite{grier2012}.
\cite{doroshenko2012} found that the rms spectrum had two prominent peaks (one is
blueshifted and the other has roughly zero velocity) during 1993 -- 1999.  The
two peaks in the rms spectrum almost disappeared in 2000 -- 2002, and rose
again in 2005 -- 2008 \citep{doroshenko2012}. A third small peak with
redshifted velocity ($\sim1500$ km~s$^{-1}$) appeared in 2005 -- 2008, which
makes the things even more complex. \cite{grier2012} found that the blueshifted
peak became  strong again in 2010, and the redshifted peak was also still
significant. It should  be noted that the blueshifted peak  always stayed at
$\sim-3000$ km~s$^{-1}$,  and did not show large velocity changes, however,
its relative intensity has changed significantly with time \citep{doroshenko2012,
grier2012}.

The velocity-resolved time lag measurements of Mrk~6 also shows  very
complex structures \citep{doroshenko2012, grier2013}.
\cite{doroshenko2012} found that the velocity-resolved lags are generally
shorter in the wings and longer in the line core, but the longest lag is
blueshifted by  $\sim-2000$ km~s$^{-1}$. They interpreted this complex
velocity-resolved lag measurement as a combination of virial motion plus
inflowing gas. The velocity-resolved lags in \cite{grier2013} are similar to the
results for 1993 -- 1995 in \cite{doroshenko2012}, but the lags increase
gradually from 2000 km~s$^{-1}$ to 7500 km~s$^{-1}$. Our velocity-resolved
lags are more similar to the results of \cite{grier2013}, that the longest lag
is located at $-2000$ km~s$^{-1}$ and the lags increase from 1500 km~s$^{-1}$
to 7000 km~s$^{-1}$. It is not easy to simply explain this complicated
velocity-resolved lag structure using a combination of virial motion and
inflowing gas suggested by \cite{doroshenko2012}. Thus, the velocity-delay map
and detailed modeling of a more complex BLR geometry and kinematics are needed to
further explore the nature of this object. It should be noted that the bin with the 
highest velocity on the red side has relatively larger uncertainty, which may be caused 
by the slight influence from the \oiii$\lambda4959$
line. We will conduct the observation in the future to check this issue.

Mrk~6 has the longest monitoring period, the highest sampling rate, and
the best S/N ratios among the 4 objects in the present paper. Although the 
continuum of Mrk~6 changes very slowly (rise slowly in the first half and 
then shows a gently falling trend, see panel c in Figure \ref{fig:light_curves}), 
which makes its ACF and CCF exceptionally broad (see also Figure \ref{fig:light_curves}), 
the uncertainties of the its time lag 
measurement produced by the FR/RSS method \citep{peterson1998, peterson2004} 
are still acceptable. But it should be pointed out that the very slow variation in the 
continuum and the corresponding broad ACF/CCF may potentially limit the smallest observable lag,
either probing different parts of the BLR or skewing the gas
distribution to respond to larger radii \citep{goad2015}. The future observation will 
investigate this possibility.

\subsubsection{SBS~1518$+$593}
\label{sec:sbs}

The time lag of SBS~1518$+$593 is mainly determined by the dip around 105 days
in the continuum light curve and the response around 125 days in the \hb\ 
light curve. The peak at $\sim$150 days in the continuum and its potential response 
at $\sim$180 in the emission-line light curve also provide constraint to the lag measurement, 
but only play a minor role because the peak in the \hb\ light curve is already near to the end
of the campaign.

The \hb\ emission line in the mean spectrum of SBS~1518$+$593 is asymmetric
toward the red wing, whereas the peak of the \hb\ in the rms spectrum is
mildly redshifted (see Figure \ref{fig:meanrms}). However, its velocity-resolved 
lag measurement is not significantly asymmetric. The longer lags in the line core and 
the shorter lags in the line wings imply that the geometry and kinematics of its BLR
tend to be virialized motion or a Keplerian disk. 

\subsection{Ongoing project}
Our primary objectives are: (1) revealing
the complex BLR physics behind AGNs with asymmetric \hb, (2)
understanding the influence of differing BLR geometry or kinematics for the
black hole mass measurement, and (3) looking for SMBH binary systems.

In particular, with
the data quality and calibration precision improving in recent years, the
velocity-resolved RM, including the velocity-resolved time lag measurements
\citep[e.g.,][]{bentz2008, bentz2009, bentz2010, denney2009, denney2010,
grier2013, du2016VI} and the analysis of the velocity-delay maps reconstructed through the
maximum entropy method \citep[e.g.,][]{horne1994, horne2004, bentz2010,
grier2013}, regularized linear inversion \citep{skielboe2015}, or Bayesian-based
dynamical modeling by Markov Chain Monte Carlo (MCMC) method
\citep[e.g.,][]{pancoast2011, pancoast2012, pancoast2014b, grier2017vdm}, was
successfully applied to more than 20 AGNs, and preliminarily revealed the geometry
and kinematics of their BLRs.   We plan to significantly add to this total and help
make a breakthrough in our understanding.

\section{Summary}
\label{sec:summary}

In this paper, we describe the MAHA project and report some results
from the first campaign.  We successfully obtained the \hb\ time lags of 4 objects
observed
from December 2016 to May 2017, and preliminarily investigated their BLR kinematics
through measuring the velocity-resolved lags. The velocity-resolved results of
3C~120, Ark~120, and Mrk~6 showed very complex structures, which were different
from the simple signatures of outflow, inflow, or virialized motion. The
velocity-resolved lag measurements of SBS~1518$+$593
showed generally shorter lags in the line wings and longer ones at the line
centers, which implied that their BLR is virialized motion or a Keplerian disk. 
The complexities of
the velocity-resolved  time lags in the AGNs with asymmetric \hb\ line
profiles clearly demonstrate the very complex geometry and kinematics of their
BLRs, and provide good opportunities to understand the physics of the BLRs in
AGNs in more details in the future.

\acknowledgments 
We thank the anonymous referee for constructive comments.
We acknowledge the support by National Key R\&D Program of
China (grants 2016YFA0400701 and 2016YFA0400702), by NSFC through grants
NSFC-11503026, -11233003, -11573026, -11773029, by the Strategic Priority 
Research Program of the Chinese Academy of Sciences Grant No. XDB23000000, and by Grant 
No. QYZDJ-SSW-SLH007 from the Key Research Program of Frontier Sciences, CAS.
This research has made use of the NASA/IPAC Extragalactic Database (NED), which is operated by the Jet Propulsion Laboratory, California Institute of Technology, under contract with the National Aeronautics and Space Administration.
We thank WIRO engineer James Weger for his invaluable assistance.

\appendix
\section{MAHA Targets}
\label{sec:maha_targets}

\begin{figure*}
\centering
\includegraphics[width=\textwidth]{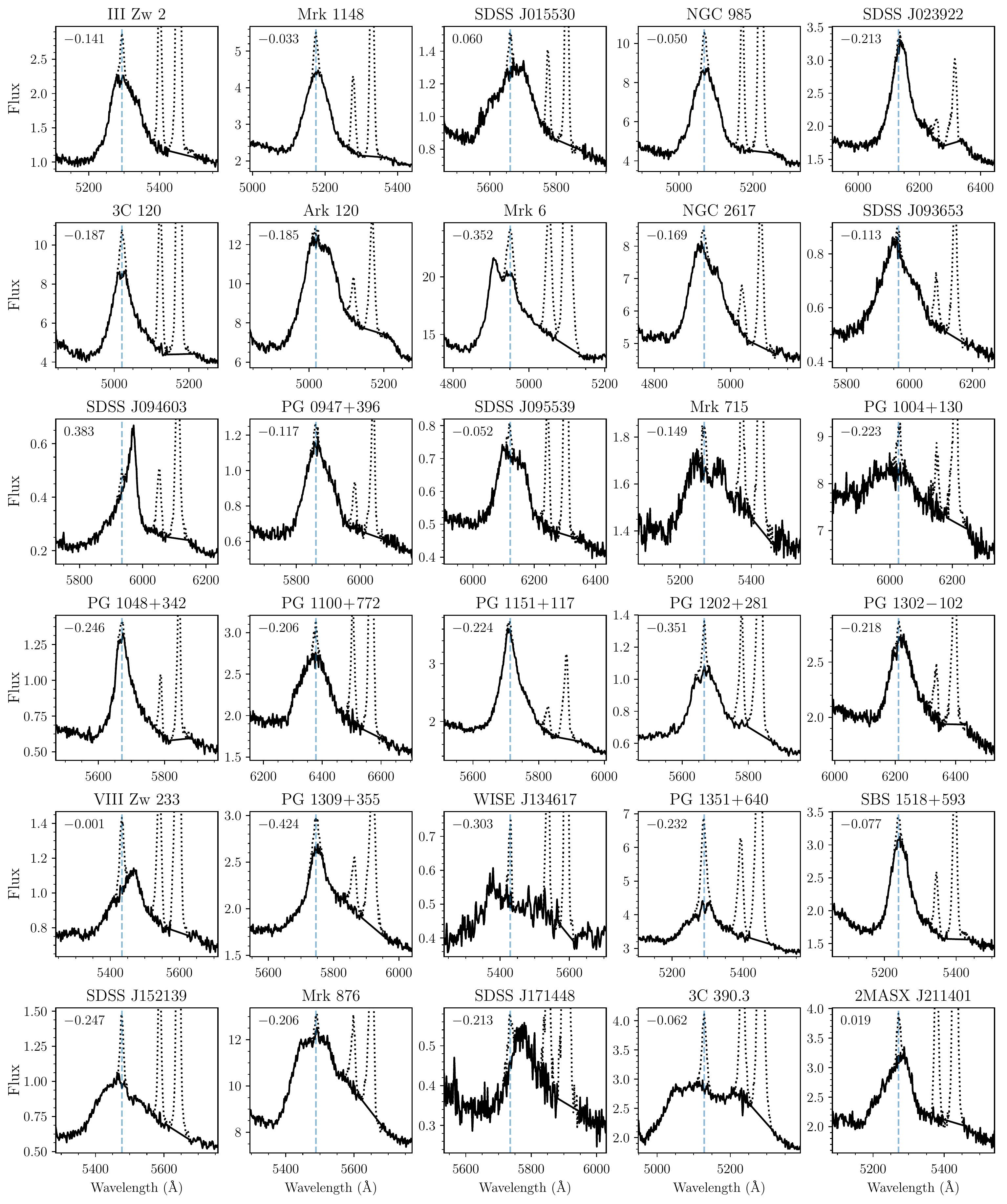}
\caption{WIRO spectra of the targets listed in Table \ref{tabmaha}. The solid lines are the narrow-line-subtracted \hb\ profiles, 
and the dotted lines are the narrow \hb, \oiii$\lambda4959,5007$ lines. The asymmetry $A$ of each object is marked in the upper-left corner.
The unit of flux is $10^{-15}~{\rm erg~s^{-1}~cm^{-2}~\AA^{-1}}$.}
\label{fig:profiles}
\end{figure*}

We present our initial suite of MAHA targets in order of Right Ascension in Table \ref{tabmaha}, 
and plot recent WIRO spectra in Figure \ref{fig:profiles}.  The WIRO spectra displayed in Figure \ref{fig:profiles} have had their narrow lines subtracted as described in the main text.

\begin{deluxetable}{lllclr}
\tablecolumns{6}
\tablecaption{MAHA Targets\label{tabmaha}}
\tablehead{
\colhead{Object}                      &
\colhead{$\alpha_{2000}$}             &
\colhead{$\delta_{2000}$}             &
\colhead{Redshift}                    &
\colhead{Mag}              &
\colhead{Asymmetry}
}
\startdata
III~Zw~2                 & 00 10 30.8 & +10 58 13 & 0.0898 & 15.4  & $-0.141$ \\
Mrk~1148                 & 00 51 54.7 & +17 25 59 & 0.0640 & 15.5V & $-0.033$ \\
SDSS~J015530.01-085704.0 & 01 55 30.0 & -08 57 04 & 0.1644 & 16.9g & $ 0.060$ \\
NGC~985                  & 02 34 37.8 & -08 47 15 & 0.0431 & 13.8  & $-0.050$ \\
SDSS~J023922.87-000119.5 & 02 39 02.9 & -00 01 29 & 0.2616 & 15.9R & $-0.213$ \\
3C~120                   & 04 33 11.1 & +05 21 16 & 0.0330 & 14.1r & $-0.187$ \\
Ark~120                  & 05 16 11.4 & -00 08 59 & 0.0327 & 15.3b & $-0.185$ \\
Mrk~6                    & 06 52 12.2 & +74 25 37 & 0.0188 & 15.0  & $-0.352$ \\          
NGC~2617                 & 08 35 38.8 & -04 05 18 & 0.0142 & 14    & $-0.169$ \\
SDSS~J093653.84+533126.8 & 09 36 53.8 & +53 31 27 & 0.2278 & 17.1g & $-0.113$ \\
SDSS~J094603.94+013923.6 & 09 46 03.9 & +01 39 24 & 0.2199 & 17.8g & $ 0.383$ \\
PG~0947+396              & 09 50 48.4 & +39 26 51 & 0.2059 & 16.3g & $-0.117$ \\
SDSS~J095539.81+453217.0 & 09 55 39.8 & +45 32 17 & 0.2594 & 16.9g & $-0.052$ \\
Mrk~715                  & 10 04 47.6 & +14 46 46 & 0.0846 & 16.6g & $-0.149$ \\
PG~1004+130              & 10 07 26.1 & +12 48 56 & 0.2406 & 15.4g & $-0.223$ \\               
PG~1048+342              & 10 51 43.9 & +33 59 27 & 0.1670 & 16.8g & $-0.246$ \\
PG~1100+772              & 11 04 13.7 & +76 58 58 & 0.3115 & 15.5B & $-0.206$ \\
PG~1151+117              & 11 53 49.3 & +11 28 30 & 0.1761 & 16.4g & $-0.224$ \\ 
PG~1202+281              & 12 04 42.1 & +27 54 12 & 0.1653 & 17.2g & $-0.351$ \\  
PG~1302$-$102            & 13 05 33.0 & -10 33 19 & 0.2784 & 14.9V & $-0.218$ \\
VIII~Zw~233              & 13 05 34.5 & +18 19 33 & 0.1180 & 16.8g & $-0.001$ \\
PG~1309+355              & 13 12 17.8 & +35 15 21 & 0.1829 & 15.8g & $-0.424$ \\
WISE~J134617.54+622045.3 & 13 46 17.5 & +62 20 45 & 0.1167 & 16.9g & $-0.303$ \\
PG~1351+640              & 13 53 15.8 & +63 45 46 & 0.0882 & 14.5V & $-0.232$ \\ 
SBS~1518+593             & 15 19 21.6 & +59 08 24 & 0.0781 & 15.8g & $-0.077$ \\
SDSS~J152139.66+033729.2 & 15 21 39.7 & +03 37 29 & 0.1265 & 17.1g & $-0.247$ \\
MRK~876                  & 16 13 57.2 & +65 43 10 & 0.1290 & 14.3V & $-0.206$ \\
SDSS~J171448.50+332738.2 & 17 14 48.5 & +33 27 38 & 0.1812 & 16.9g & $-0.213$ \\
3C~390.3                 & 18 42 09.0 & +79 46 17 & 0.0561 & 14.37 & $-0.062$ \\
2MASX~J21140128+8204483  & 21 14 01.2 & +82 04 48 & 0.0840 & 15.7  & $ 0.019$ \\
\enddata
\tablecomments{
Positions, redshifts, and magnitudes are from
the NASA/IPAC Extragalactic Database (NED), except for the magnitude of 
III~Zw~2, for which NED has no entry.  We have provided our own estimate of 15.4.
We have calculated asymmetries using measurements of new single-epoch
WIRO spectra shown in Figure \ref{fig:profiles} and the formula of \cite{DeRobertis1985}.
}
\end{deluxetable}

Below are notes on individual objects in alphanumeric order by name.

{\em 2MASX~J21140128+8204483}.  Also known as S5 2116+81, the H$\beta$ profile displays a significant blue asymmetry.

{\em 3C~120.}  One of the objects featured in this paper and possessing an H$\beta$ line with a red asymmetric tail.  3C 120 has been previously reverberation mapped several times, with recent results published by \cite{grier2012}.

{\em 3C~390.3.}  Others have previously reverberation mapped this object, which displays
double-peaked emission lines
 \citep[e.g.,][]{shapovalova2010,dietrich2012,sergeev2017}.
 3C~390.3  once displayed quasi-periodic emission-line profile changes suggestive of a binary \citep{gaskell1996}, but later deviated from the predicted pattern \citep{eracleous1997}.

{\em III~Zw~2}. This object has an H$\beta$ profile with an extreme red asymmetry.

{\em Ark~120}. One of the objects featured in this paper, displaying a broad and complex double-peaked H$\beta$ profile.  Previous RM results exist in the literature \citep[e.g.,][]{peterson1998,doroshenko2008}.

{\em Mrk~6}. One of the objects featured in this paper and possessing a highly blueshifted peak and a red tail, probably also double-peaked at the present epoch.  There has been high-quality RM at previous epochs \citep{doroshenko2012, grier2012}.

{\em Mrk~715}. Also known as SDSS~J100447.61+144645.6, this object has a double-peaked H$\beta$ line profile and a long tail to the red.

{\em Mrk~876}. The current H$\beta$ profile is suggestive of a double-peaked profile with red asymmetry.  In the past \citep{DeRobertis1985}, the blueshifted peak was significantly stronger, reminiscent of Mrk 6.

{\em Mrk~1148}. This object has an H$\beta$ profile with a mild red asymmetry.

{\em NGC~985}. This object has an H$\beta$ profile with a red asymmetry.

{\em NGC~2617}. Recently \citet{fausnaugh2017} published RM of this object.  Interestingly, their work covering the 2014 epoch shows the H$\beta$ profile has a significant bump on the {\em blue} side of the profile, while WIRO spectra in 2017 and 2018 show a bump on the {\em red} side of the profile.

{\em PG~0947+396}. The H$\beta$ profile shows a red asymmetry.

{\em PG~1004+130}.  This object is somewhat luminous ($> 10^{45}$ ergs s$^{-1}$) and a likely radio-loud broad absorption line quasar \citep{wills1999}.  The H$\beta$ line has a red asymmetry.

{\em PG~1048+342}. This PG quasar has a red asymmetric H$\beta$ profile.

{\em PG~1100+772}.   This luminous PG quasar has a H$\beta$ profile with a blue bump but a red tail.

{\em PG~1151+117}.  The H$\beta$ profile shows a red asymmetry.  

{\em PG~1202+281}. Also known as GQ COM, this object has an H$\beta$ profile with a red asymmetry.

{\em PG~1302-102}.  This luminous radio-loud PG quasar has a red asymmetric H$\beta$ profile.  More notably, \citet{graham2015} find a $\sim$5 year periodicity suggesting a binary nature, although  \citet{liu2018} suggest that the periodicity may have vanished.  Time will tell.

{\em PG~1309+355}. This PG quasar has a red asymmetric H$\beta$ profile, and appears particular well fit by two Gaussians suggesting two components.

{\em PG~1351+640}. This object has an H$\beta$ profile with a bump on the blue side but also a long red wing.  The bump seems to have weakened compared to the spectrum shown by \citet{boroson1992} observed over 25 years previously.

{\em SBS~1518+593}. One of the objects featured in this paper, featuring a mild red asymmetry at the present epoch.

{\em SDSS~J015530.01-085704.0}. The H$\beta$ line has a significantly redshifted peak
along with the customary associated blue asymmetry \citep{eracleous2012}.

{\em SDSS~J023922.87-000119.5}. The H$\beta$ profile shows a red asymmetry.

{\em SDSS~J093653.84+533126.8}. In SDSS spectra, this object has an H$\beta$ profile very similar to that of SDSS~J094603.94+013823.6, showing a redshifted peak and blue asymmetry.  Since then, a redshifted component has weakened dramatically leaving an emission line that is much more symmetric \citep{runnoe2017}.  All that is left now of that strong component is a weak, redshifted bump.  

{\em SDSS~J094603.94+013823.6}. The H$\beta$ line has a significantly redshifted peak
along with the customary associated blue asymmetry \citep{eracleous2012}.  
The shifting profile is still consistent with expectations for a CB-SMBH \citep{runnoe2017}.

{\em SDSS~J095539.81+453217.0}. The H$\beta$ line has a flat top with a blueshifted peak and a red asymmetry.

{\em SDSS~J152139.66+033729.2}.  This object possesses an H$\beta$ line with a red asymmetry. 

{\em SDSS~J171448.50+332738.2}.  This object possesses an unusual H$\beta$ line with a redshifted top and red asymmetric wing.

{\em VIII~Zw~233}.  The H$\beta$ line has a significantly redshifted peak.

{\em WISE~J134617.54+622045.3}. The H$\beta$ line has a significantly blueshifted peak along with the customary red asymmetry \citep{eracleous2012}.  The profile is reminiscent of Mrk~6.

\section{Velocity-resolved time lags based on mean spectra}
\label{sec:v_resolved_lag_mean}

\begin{figure*}
\centering
\includegraphics[width=\textwidth]{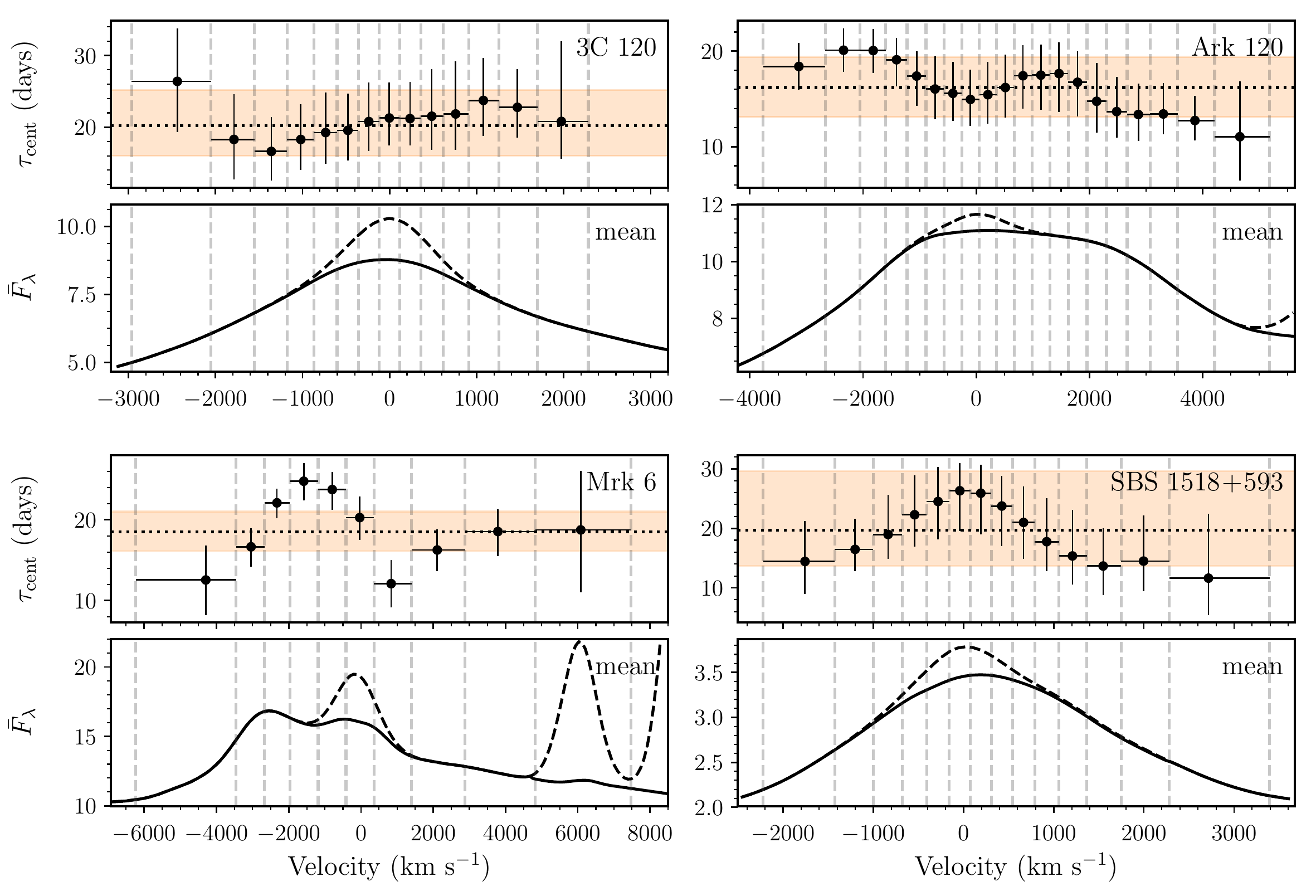}
\caption{Rest-frame velocity-resolved time lags and the corresponding mean spectra. The upper panel in each plot shows
the centroid lags at different velocities, and the lower panel is the mean spectrum in unit of ${10^{-15}\ \rm erg\ s^{-1}\
cm^{-2}\ \AA^{-1}}$. The name of the object is marked in the upper-right corner of each plot. The vertical dashed lines are the 
edges of velocity bins. The horizontal dotted lines and the orange color are the average time lags and the uncertainties 
in Table \ref{tab:lags}. The black dashed lines in the lower panels are the narrow emission lines.}
\label{fig:v_resolved_lag_mean}
\end{figure*}

The velocity-resolved lags in Section \ref{sec:velocity_resolved_lags} are
measured based on the velocity bins with equal flux in the rms spectrum,
where the emission line in each bin have the same level of variation but not
the physical flux. As a further test, we divided the \hb\ lines into the
velocity bins, each of which having the same \hb\ fluxes in the narrow-line-subtracted 
mean spectra, and measured the velocity-resolved lags again.
Figure \ref{fig:v_resolved_lag_mean} demonstrates the velocity-resolved lag
measurements based on the mean spectra. Comparing with the rms-based results, the bins in high velocities
become relatively narrower and the bins in low velocities become wider, because the \hb\ profiles in the 
mean spectra are broader than those in the rms spectra for these objects (see Table \ref{tab:fwhm_mbh}). 
For Mrk~6, the red wing is located beneath the \oiii$\lambda4959$ 
(see Figure \ref{fig:v_resolved_lag_mean}). We reduced 
the number of bins (compared with the rms-based result of Mrk~6) to make the bins wider that the highest velocity bin 
can exactly cover the \oiii$\lambda4959$
in order to avoid the potential influence from the strong \oiii. For the other 3 objects, the number of bins and the bluest and 
reddest velocity limits are the same as the rms-based results (Figure \ref{fig:v_resolved_lags}). 
In general, the results are almost the
same as the velocity-resolved lags in Section
\ref{sec:velocity_resolved_lags}, which means the velocity-resolved 
analysis in this work is robust.

\end{document}